\documentclass[aps,pra,showpacs,superscriptaddress,tightenlines,twocolumn,lengthcheck]{revtex4-1}
\usepackage{diagbox}
\usepackage{multirow}
\usepackage{amsmath}
\usepackage{amsfonts}
\usepackage{graphicx}
\usepackage{epsfig}
\usepackage{color}
\usepackage{txfonts}
\usepackage[colorlinks,citecolor=blue]{hyperref}
\usepackage{diagbox}
\usepackage{mathrsfs}
\usepackage{makecell}
\usepackage{xcolor}
\usepackage{amsfonts}
\usepackage{bbm}
\usepackage{dsfont}
\usepackage[T1]{fontenc}

\begin{document}
	
		\title{Harnessing dark states: coherent control in coupled cavity-Rydberg-atom systems}
		\author{Ying-Zhi Li}
		\email{These authors contributed equally to this work.}
		\affiliation{Key Laboratory of Low-Dimensional Quantum Structures and Quantum Control of Ministry of Education, Key Laboratory for Matter Microstructure and Function of Hunan Province, Department of Physics and Synergetic Innovation Center for Quantum Effects and Applications, Hunan Normal University, Changsha 410081, China}
		\affiliation{Hunan Research Center of the Basic Discipline for Quantum Effects and Quantum Technologies, Hunan Normal University, Changsha 410081, China}
		\author{Xuan Zhao}
		\email{These authors contributed equally to this work.}
		\affiliation{Key Laboratory of Low-Dimensional Quantum Structures and Quantum Control of Ministry of Education, Key Laboratory for Matter Microstructure and Function of Hunan Province, Department of Physics and Synergetic Innovation Center for Quantum Effects and Applications, Hunan Normal University, Changsha 410081, China}
		\affiliation{Hunan Research Center of the Basic Discipline for Quantum Effects and Quantum Technologies, Hunan Normal University, Changsha 410081, China}
		\author{Le-Man Kuang}
		\affiliation{Key Laboratory of Low-Dimensional Quantum Structures and Quantum Control of Ministry of Education, Key Laboratory for Matter Microstructure and Function of Hunan Province, Department of Physics and Synergetic Innovation Center for Quantum Effects and Applications, Hunan Normal University, Changsha 410081, China}
		\affiliation{Hunan Research Center of the Basic Discipline for Quantum Effects and Quantum Technologies, Hunan Normal University, Changsha 410081, China}
		\author{Jie-Qiao Liao}
		\email{Contact author: jqliao@hunnu.edu.cn}
		\affiliation{Key Laboratory of Low-Dimensional Quantum Structures and Quantum Control of Ministry of Education, Key Laboratory for Matter Microstructure and Function of Hunan Province, Department of Physics and Synergetic Innovation Center for Quantum Effects and Applications, Hunan Normal University, Changsha 410081, China}
		\affiliation{Hunan Research Center of the Basic Discipline for Quantum Effects and Quantum Technologies, Hunan Normal University, Changsha 410081, China}
		\affiliation{Institute of Interdisciplinary Studies, Hunan Normal University, Changsha 410081, China}

	\begin{abstract}
			
		The dark-state effect, caused by destructive interference, not only is an important fundamental research topic in atomic physics and quantum optics, but also has wide potential application in quantum physics and quantum information science. Using the arrowhead-matrix method, here we study the dark-state effect in a coupled cavity-Rydberg-atom system, in which $N$ Rydberg atoms with the dipole-dipole interactions are coupled to a single-mode cavity field. We obtain the numbers and form of the dark states in certain excitation-number subspaces for the two-, three-, and four-atom cases, as well as in the single-excitation subspace for a general $N$-atom case. We also suggest to characterize the dark states by inspecting the populations of some specific quantum states, which can be detected in experiments. Furthermore, we analyze the dark-state effect in a realistic case, where both the atomic dipole-dipole interaction strengths and the atom-cavity-field coupling strengths depend on the position of the atoms. Our findings pave the way for studying dark-state physics and applications in the cavity-Rydberg-atom platform.

	\end{abstract}
	
	\date{\today}
	\maketitle
	
	\section{Introduction}
	
	The Rydberg-atom systems~\cite{ryd-re1,ryd-re2,ryd-re3,ryd-re4,ryd-re5} have become an important physical platform for exploring quantum simulation~\cite{ryd-QS1,ryd-QS2} and quantum information~\cite{ryd-QI-cat,ryd-QI-sci,ryd-QI-li}, owing to the strong long-range dipole-dipole interaction and highly flexible geometry facilitated by individual optical-tweezer trapping techniques~\cite{flex2006}. In particular, Rydberg-atom systems have been suggested to simulate a variety of many-body quantum models~\cite{ryd-man-body-na,ryd-many-body-sci} and strongly correlated phenomena, such as magnetism and dynamics in quantum spin models~\cite{spin2017x,spin2018}, symmetry protected topological phase~\cite{phase2019}, emergent gauge field~\cite{gauge2020a,gauge2020f,gauge2024}, coherent excitation transfer~\cite{excit2015,excit2024}, and many-body localization~\cite{local2017,local2021}. To further integrate the unique advantages of the Rydberg-atom systems and optical cavities, much recent attention has been paid to the coupled cavity-Rydberg-atom systems~\cite{cavity-ryd-atom1,cavity-ryd-atom2,cavity-ryd-atom3,cavity-ryd-atom4,cavity-ryd-atom5,cavity-ryd-atom6,cavity-ryd-atom7,cavity-atom-y}, which provide a versatile physical platform for exploring novel physical effects and quantum information applications based on quantum light fields and Rydberg atoms.
	
	Due to the high-dimensional state space of the cavity field and its role as a mediator coupling all atoms, the coupled cavity-Rydberg-atom systems exhibit rich energy level structures and transitions, which provide the physical conditions for the occurrence of novel quantum interference phenomena. The dark-state effect~\cite{experiment1976}, originated from the destructive quantum interference, is a significant physical effect in atomic physics and quantum optics~\cite{scully}. Due to its unique properties, the dark-state effect has wide application in both fundamental quantum physics~\cite{ds-fund-2009,ds-fund-2013,ds-fund-2017,ds-fund-2022-1,ds-fund-2022-2,zhao2026} and modern quantum science and technology~\cite{science-2020,science-2021,science-2022-1,science-2022-2}. A large number of physical effects associated with the dark states have been extensively studied, such as coherent population trapping~\cite{cpt1978,cpt1982,cpt1988,cpt1996,cpt1998}, electromagnetically induced transparency~\cite{eit1991,eit1996,eit1997,eitds2000,eit2005}, and stimulated Raman adiabatic passage~\cite{stirap1989,stirap2015,stirap2017}. Recently, much attention has been paid to the dark states in various physical systems, such as cavity-QED~\cite{cavity-QED1,cavity-QED2,cavity-QED3,cavity-QED4,cavity-QED5,cavity-QED6,cavity-QED7}, circuit-QED systems~\cite{circuit-QED1,circuit-QED2,circuit-QED3}, and waveguide-QED systems~\cite{waveguide-QED1,waveguide-QED2,waveguide-QED3,waveguide-QED4}, creating a new frontier for implementation of quantum information processing using dark states. Furthermore, the concept of dark states has been extended to the dark modes in coupled atom-field systems~\cite{dm-atom1,dm-atom2} and coupled bosonic-mode systems~\cite{dm-bose-Genes,dm-bose-dch,dm-bose-wyd,dm-bose-tl,dm-bose-ldg1,dm-bose-ldg2,dm-bose-ldg3,dm-bose-ldg4}. Specifically, a general method, the arrowhead-matrix method, for determining the number and form of orthogonal dark modes in bosonic networks has been proposed~\cite{huang2023}. This method has recently been generalized to study the dark-state effect in the Fock-state lattices~\cite{zhao2025} and arbitrary multilevel quantum systems~\cite{zhao2026}. Nevertheless, the dark-state effect in a coupled cavity-atom systems with interatomic interactions, particularly the dipole-dipole interaction of the Rydberg atoms, remains unexplored.
	
	In this work, we consider a coupled cavity-Rydberg-atom system and analyze the dark-state effect using the arrowhead-matrix method~\cite{huang2023}. Firstly, we define the upper states (the states associated with the cavity excited states) and the lower states (the states associated with the cavity vacuum state), then the basis states of the system can be divided into two sub-components. By defining the basis vectors in a given excitation-number subspace, the Hamiltonian of the system can be expressed as a block matrix, where two submatrices on the diagonal correspond to the upper- and lower-state components, and the remaining submatrices describe the couplings between the two sub-components. By diagonalizing the lower-state submatrix, the Hamiltonian can be expressed as an arrowhead matrix. Then the number and form of the dark states can be obtained with the arrowhead-matrix method~\cite{huang2023}. Concretely, we obtain the numbers and form of the dark states in certain excitation-number subspaces for the two-, three-, four-atom cases. Furthermore, we obtain the number and form of the dark states in the single-excitation subspace of a general coupled cavity-$N$-atom system. Furthermore, we find that the dark-state effect in the system can be characterized by inspecting the population of some specific states, which can be detected in experiments. We also analyze the dark-state effect in a realistic case, where both the atomic dipole-dipole interaction strengths and the atom-cavity-field coupling strengths depend on the position of the atoms. We find the detailed parameter conditions for the appearance of the dark states in the system.
	
	The rest of this work is organized as follows. In Sec.~\ref{sec2}, we introduce the coupled cavity-Rydberg-atom system and present the arrowhead-matrix method for analyzing the dark-state effect in this system. In Secs.~\ref{sec3}, ~\ref{sec4}, and~\ref{sec5}, we derive the numbers and form of the dark states in different-excitation subspaces corresponding to the two-, three-, four-atom cases. In particular, we characterize the dark-state effect from the population of some selected states. In Sec.~\ref{sec6}, we obtain the dark states in the single-excitation subspace for the $N$-atom case. In Sec.~\ref{sec7}, we study the dark-state effect when both the atomic dipole-dipole interaction strengths and atom-cavity-field coupling strengths depend on the position of the atoms. Finally, we conclude this work in Sec.~\ref{sec8}.

	\section{physical model and the arrowhead-matrix method}\label{sec2}

\begin{figure}[t!]
		\centering\includegraphics[width=0.48\textwidth]{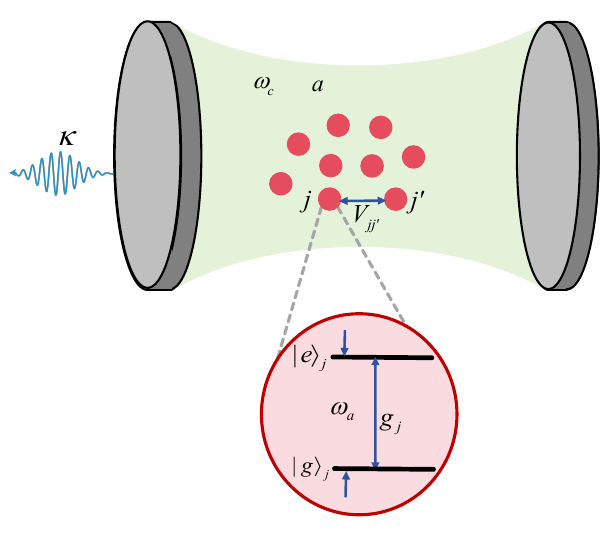}
		\caption{Schematic of the coupled cavity-Rydberg-atom system composed of a cavity with the resonance frequency  $\omega_{c}$ and $N$ Rydberg atoms with the energy separation  $\omega_{a}$ (between the excited state $\left\vert e\right\rangle $ and ground state $\left\vert g\right\rangle $). The variable $V_{jj'}$ denotes the dipole-dipole interaction strength between the $j$th and $j'$th atoms, and $g_j$ describes the Jaynes-Cummings-type coupling between the $j$th atom and the cavity field. The parameter $\kappa$ is the decay rate of the cavity mode.}
		\label{model}
	\end{figure}

	 We consider a coupled cavity-Rydberg-atom system in which $N$ Rydberg atoms are coupled to a single-mode field in the cavity through the Tavis-Cummings-type interactions~\cite{Tc}, and these atoms are coupled to each other through the dipole-dipole interactions (see Fig.~\ref{model}). The Hamiltonian of this system reads
	(with $\hbar =1$)~\cite{cavity-atom-y}%
	\begin{eqnarray}
		\hat{H}_{[N]}^{\text{sys}} &=&\frac{\omega _{a}}{2}\sum_{j=1}^{N}\hat{\sigma}
		_{j}^{z}+\sum_{j<j^{\prime }}^{N}V_{jj^{\prime }}(\hat{\sigma}_{j}^{+}\hat{\sigma} _{j^{\prime }}^{-}+\hat{\sigma}_{j^{\prime }}^{+}\hat{\sigma}
		_{j}^{-})   \notag \\
		&&+\omega _{c}\hat{a}^{\dag }\hat{a}+\sum_{j=1}^{N}g_{j}( \hat{a}^{\dag }\hat{\sigma}
		_{j}^{-}+\hat{\sigma} _{j}^{+}\hat{a}) ,
		\label{Hamtotal}
	\end{eqnarray}
	where $\omega_{a}$ is the energy separation between the excited state $\left\vert e\right\rangle _{j}$ and ground state $\left\vert g\right\rangle _{j}$ of the $j$th ($j=1,2,\dots,N$) atom, which is described by the Pauli operators  $\hat{\sigma}_{j}^{x}=\left| e \right\rangle_{jj}\!\left\langle g \right| +\left\vert g\right\rangle_{jj}\! \left\langle
	e\right\vert $, $\hat{\sigma}_{j}^{y}=i(\left\vert g\right\rangle_{jj}\! \left\langle
	e\right\vert -\left\vert e\right\rangle_{jj}\! \left\langle g\right\vert )$, and $\hat{\sigma}_{j}^{z}=\left\vert	e\right\rangle_{jj} \!\left\langle e\right\vert -\left\vert g\right\rangle_{jj}\!
	\left\langle g\right\vert $. The raising and lowering operators of the $j$th atom are defined by $\hat{\sigma} _{j}^{\pm}=(\hat{\sigma}_{j}^{x}\pm i\hat{\sigma}_{j}^{y})/2$. The variable $V_{jj^{\prime }}$ is the dipole-dipole interaction strength between the $j$th and $j'$th atoms. The parameter $\omega_{c}$ is the resonance frequency of the field mode described by the annihilation (creation) operator $\hat{a}$ $(\hat{a}^{\dag })$, and $g_{j}$ is the Jaynes-Cummings-type coupling~\cite{Jc} strength between the $j$th atom and the field mode. Note that the subscript \textquotedblleft $\left[ N\right] $\textquotedblright\ in Eq.~(\ref{Hamtotal}) is introduced to mark the number of the involved atoms.
	
     For better analyzing the dark-state effect, we work in a rotating frame
     defined by the unitary operator $\hat{U}=\exp [ -i\omega _{c}t( \hat{a}^{\dag}\hat{a}+\sum_{j=1}^{N}\hat{\sigma} _{j}^{z}/2) ] ,$ then the Hamiltonian in
     Eq.~(\ref{Hamtotal}) becomes%
     \begin{equation}
     	\hat{H}_{[N]} =\frac{\Delta _{a}}{2}\sum_{j=1}^{N}\hat{\sigma}
     	_{j}^{z}+\sum_{j<j^{\prime }}^{N}V_{jj'}( \hat{\sigma} _{j}^{+}\hat{\sigma} _{j^{\prime
     	}}^{-}+\hat{\sigma} _{j}^{-}\hat{\sigma} _{j^{\prime }}^{+})+\sum_{j=1}^{N}g_{j}( \hat{a}^{\dag }\hat{\sigma} _{j}^{-}+\hat{\sigma}_{j}^{+}\hat{a})   , \label{Hn}
     \end{equation}%
     where we introduce the detuning $\ \Delta _{a}=\omega _{a}-\omega _{c}$ between the atomic energy separation $\omega _{a}$ and the cavity field frequency $\omega _{c}$. 
     
     In the closed-system case, the total excitation number operator $\mathrm{\hat{N}}_{[N]}=\hat{a}^{\dag }\hat{a}+\sum_{j=1}^{N}\hat{\sigma} _{j}^{+}\hat{\sigma}
     _{j}^{-}$ is a conserved quantity because of $[ \hat{H}_{[N]},\mathrm{\hat{N}}_{[N]}%
     ] =0.$ As a result, below we study the dark-state effect in
     different excitation-number subspaces. In particular, we will consider two different cases corresponding to $ n< N $ and $ n\geq N $. This is because the results concerning the dark states are different for these two cases.
     
     When the excitation number $n$ is less than the atom number $N$, i.e., $n<N$, the basis states in the $n$-excitation subspace can be arranged in a descending order of the cavity field excitation number and the atomic excitation in Table~\ref{table}.
     \begin{table}[t!]
     	\caption{The basis states in the $n$-excitation subspace of the coupled cavity-$N$-atom system $( n< N) $.}
     	\centering
     	\renewcommand{\arraystretch}{2.3}
     	\begin{tabular}{|c|c|c|}
     		\hline
     		\makecell{The number of\\photons in the cavity} & 
     		\makecell{The form of\\the basis states} & 
     		\makecell{The number of\\basis states} \\ \hline
     		$n$ & $\left| n,g_1,\dots,g_N\right\rangle$ & $C_{N}^{0}$ \\ \hline
     		$n-1$ & $\begin{aligned}
     			&\left| n-1,e_1,g_2,\dots,g_N\right\rangle \\
     			&\left| n-1,g_1,e_2,\dots,g_N\right\rangle \\
     			&\qquad\dots \\  
     			&\left| n-1,g_1,g_2,\dots,e_N\right\rangle
     		\end{aligned}$ & ${C_{N}^{1}}$ \\ \hline
     		\dots & \dots & \dots \\ \hline
     		$1$ & $\begin{aligned}
     			&\left| 1,e_1,\dots,e_{n-1},\dots,g_N\right\rangle \\
     			&\left| 1,e_1,\dots,g_{n-1},e_n,\dots,g_N\right\rangle \\
     			&\qquad\dots \\  
     			&\left| 1,g_1,\dots,e_{N-n+2},\dots,e_N\right\rangle
     		\end{aligned}$ & ${C_{N}^{n-1}}$ \\ \hline
     		$0$ & $\begin{aligned}
     			&\left| 0,e_1,\dots,e_n,\dots,g_N\right\rangle \\
     			&\left| 0,g_1,e_2,\dots,e_{n+1},\dots,g_N\right\rangle \\
     			&\qquad\dots \\  
     			&\left| 0,g_1,\dots,e_{N-n+1},\dots,e_N\right\rangle
     		\end{aligned}$ & ${C_{N}^{n}}$ \\ \hline
     	\end{tabular}
     	\label{table}
     \end{table}
     When all $n$ excitations are stored in the cavity field, all these $N$ atoms will be their ground states. In this case, there is only one basis state, and the number of the basis state can be expressed as $C_{N}^{0}$. When the cavity mode contains $n-1$ excitations, the remaining one excitation will be stored in these $N$ atoms. In this case, the distribution of the single excitation in these $N$ atoms can be described by a classical permutation and combination problem. Therefore, there are $C_{N}^{1}$ possible arrangements when $n-1$ excitations are in the cavity field. Similarly, we can obtain the number of the basis states for other cases. In particular, when there is one excitation in the cavity field, then other $n-1$ excitations will be stored in these $N$ atoms. Therefore, there are $C_{N}^{n-1}$ distributions for this case because each atom can possess at most one excitation. When the cavity field is in vacuum, then all these $n$ excitations are in these $N$ atoms. In this case, the number of the basis state is $C_{N}^{n}$. Based on the above analyses, we know that the number of the basis states is $C_{N}^{0}+C_{N}^{1}+C_{N}^{2}+\dots+C_{N}^{n-1}+C_{N}^{n}$ in the $n$-excitation subspace. 
     
     In the present coupled cavity-atom systems, the dark states refer to those states decoupled from the cavity field. Therefore, depending on whether the cavity field is in its excited state $\left\vert n\right\rangle (n>0 )$ or the vacuum state $\left\vert 0\right\rangle $, we
     can define these basis states as the upper and lower states, respectively. Furthermore, all these basis states of the system can be divided into two sub-components: the upper-state component and the lower-state component. In the $n$-excitation subspace, the number of the upper and lower states are $N_{u}=C_{N}^{0}+C_{N}^{1}+C_{N}^{2}+\dots+C_{N}^{n-1}$ and $N_{l}=C_{N}^{n}$, respectively. We denote these $N_{u}$ upper states as $\{\vert u_{1}\rangle ,\vert u_{2}\rangle,\dots,\vert u_{N_{u}}\rangle \}$ and $N_{l}$ lower states as $\{\vert l_{1}\rangle ,\vert l_{2}\rangle,\dots,\vert l_{N_{l}}\rangle \}$, then the basis vectors for these upper and lower states can be defined as~\cite{zhao2026}
     \begin{subequations}
     	\begin{align}
     		\left\vert u_{1}\right\rangle &= \left( 1_{1},0,\dots,0,0,0,\dots,0\right) ^{T}, \\
     		\left\vert u_{2}\right\rangle &= \left( 0,1_{2},\dots,0,0,0,\dots,0\right) ^{T}, \\
     		\notag 
     		\dots\\
     		\left\vert u_{N_{u}}\right\rangle &= \left( 0,0,\dots,1_{N_{u}},0,0,\dots,0\right)^{T},\\
     		\left\vert l_{1}\right\rangle &= \left( 0,0,\dots,0,1_{N_{u}+1},0,\dots,0\right)^{T},\\
     		\left\vert l_{2}\right\rangle &= \left( 0,0,\dots,0,0,1_{N_{u}+2},\dots,0\right)^{T},\\
     		\notag 
     		\dots\\
     		\left\vert l_{N_{l}}\right\rangle &= \left(0,0,\dots,0,0,0,\dots,1_{N_{u}+N_{l}}\right) ^{T},
     	\end{align}
     \end{subequations}
     where the subscript $s$ [for $s=1$-($N_{u}+N_{l}$)] of the element \textquotedblleft $1_s$\textquotedblright\ in these basis vectors is introduced to denote its position in the vector, and the superscript \textquotedblleft $T$\textquotedblright\ denotes the matrix transpose.
     In this representation, the Hamiltonian of the system restricted within the $n$-excitation subspace can be expressed as a matrix%
     \begin{equation}
     	H_{[N]}^{\left( n\right) }=\left(
     	\begin{array}{c|c}
     		\mathbf{U}_{\left[ N\right] }^{\left( n\right) } & \mathbf{C}_{\left[ N%
     			\right] }^{\left( n\right) } \\ \hline
     		\left( \mathbf{C}_{\left[ N\right] }^{\left( n\right) }\right) ^{\dag } &
     		\mathbf{L}_{\left[ N\right] }^{\left( n\right) }%
     	\end{array}%
     	\right) , \label{HNn}
     \end{equation}%
     where $\mathbf{U}_{\left[ N\right] }^{\left( n\right) },$ $\mathbf{L}_{\left[
     	N\right] }^{\left( n\right) },$ and $\mathbf{C}_{\left[ N\right] }^{\left(
     	n\right) }$ are the submatrices associated with the upper-state component,
     lower-state component, and the couplings between the two sub-components of
     states, respectively. Note that the superscript
     \textquotedblleft $\left( n\right) $\textquotedblright\ in these matrices is introduced to indicate the excitation number related to the subspace.
     
     When the excitation number is equal to the atom number $n=N$, there is one lower state with all $n$ excitations stored in the $N$ atoms. With the further increase of the excitation number $n$, the excess excitations (from $N+1$ to $n$) can only be stored in the cavity field. Therefore, there is no lower state when $n>N$. As a result, for the case $n\geq N$, there is no dark state because the necessary interference channels are absent in the lower states. In the following we will only consider the case $n<N$.
     
     For the case $n<N$, all the matrix elements in Eq.~(\ref{HNn}) can be calculated based
     on the given basis vectors and the Hamiltonian in Eq.~(\ref{Hn}). Since there exist dipole-dipole interactions among atoms, the submatrices $\mathbf{U}_{\left[ N\right] }^{\left( n\right)
     } $ and $\mathbf{L}_{\left[ N\right] }^{\left( n\right)
     } $ are off-diagonal. We know that the couplings among these upper states will not change the number of the dark states~\cite{huang2023}. In this work, we only diagonalize the lower-state submatrix to study the dark states for simplicity. By introducing the unitary matrix $\mathbf{S}_{l}$, the lower-state submatrix $\mathbf{L}_{\left[ N\right] }^{\left( n\right) }$ can be diagonalized as $\mathbf{\tilde{L}}_{\left[ N\right] }^{\left( n\right) }=\mathbf{S}_{l}\mathbf{L}_{\left[ N\right] }^{\left( n\right) }\mathbf{S}_{l}^{\dag }$, and the corresponding coupling matrix $\mathbf{C}_{\left[ N\right] }^{\left( n\right) }$ becomes $\mathbf{\tilde{C}}_{\left[ N\right] }^{\left( n\right) }=\mathbf{C}_{\left[ N%
     \right] }^{\left( n\right) }\mathbf{S}_{l}^{\dag }$. Then the Hamiltonian can be expressed as an arrowhead matrix~\cite{huang2023}
     \begin{equation}
     	\tilde{H}_{[N]}^{\left( n\right) }=\left(
     	\begin{array}{c|c}
     		\mathbf{U}_{\left[ N\right] }^{\left( n\right) } & \mathbf{\tilde{C}}_{\left[ N%
     			\right] }^{\left( n\right) } \\ \hline
     		\left( \mathbf{\tilde{C}}_{\left[ N\right] }^{\left( n\right) }\right) ^{\dag } &
     		\mathbf{\tilde{L}}_{\left[ N\right] }^{\left( n\right) }%
     	\end{array}%
     	\right) , \label{HNnd}
     \end{equation}
     with the diagonal submatrix $\mathbf{\tilde{L}}_{\left[ N\right] }^{\left( n\right) }$.
     Here, we introduce the dressed lower states as 
     $\{\vert L_{1}\rangle ,\vert L_{2}\rangle,\dots,\vert L_{N_{l}}\rangle \}$, which are the eigenstates of the lower-state submatrix $\mathbf{\tilde{L}}_{\left[ N\right] }^{\left( n\right) }$. According to the dark-state theorems~\cite{huang2023,zhao2026}, the number and form of the dark states can be obtained by analyzing the arrowhead matrix in Eq.~(\ref{HNnd}).
     
     A further task for exploring the dark-state effect is the characterization of the dark state. Typically, we can witness the existence of the dark state by analyzing its dynamics. 
     For the open-system case, we adopt the quantum master equation to describe the evolution of the system. For typical coupled cavity-Rydberg-atom systems, the decay rate of the atom is much smaller than that of the cavity field. Therefore, we can neglect the atomic dissipation when we characterize the dark states in the open-system case. In this case, the environment of the cavity field can be modeled by a vacuum bath. Note that since the dark states only involve atomic excitations, then the atomic dissipation will eventually cause the dark states to decay to the ground state of the system in the long-time limit. Therefore, the dynamical dark-state characterization only works within the duration $1/\kappa < t\ll 1/\gamma $, where $\kappa $ and $\gamma $ are the decay rates of the cavity field and atoms, respectively. Note that here we consider the same dissipation for all atoms. In this case, the dynamics of the coupled cavity-atom system can be approximately described by the Lindblad quantum master equation~\cite{scully}%
     \begin{equation}
     	\dot{\hat{\rho}}=i\left[ \hat{\rho} ,\hat{H}_{[N]}\right] +\mathcal{L}_{c}\left( \hat{\rho}\right).\label{master}
     \end{equation}%
     Here, we introduce the Lindblad superoperator $\mathcal{L}_{c}\left( \hat{\rho} \right)$ to describe the dissipation of the cavity field, and it is given by%
     \begin{subequations}
     	\begin{align}
     		\mathcal{L}_{c}\left( \hat{\rho} \right) =&\frac{\kappa }{2}( 2\hat{a}\hat{\rho} \hat{a}^{\dag
     		}-\hat{a}^{\dag }\hat{a}\hat{\rho} -\hat{\rho} \hat{a}^{\dag }\hat{a})  ,
     	\end{align}%
     \end{subequations}
     where $\kappa$ is the decay rate of the cavity field.
     
     For characterization of the dark-state effect, we investigate the dynamics of the system by preparing the system in proper initial states. To create the expected initial state of the system, we introduce the cavity-field driving and the atomic drivings, which are described by the driving Hamiltonians~\cite{walls}
     \begin{subequations}\label{qudong}
     	\begin{align}
     		H_{cd} &= \Omega _{cd}( a^{\dag }e^{-i\omega _{cd}t}+ae^{i\omega _{cd}t}),\label{qudong-c} \\
     		H_{ad} &= \Omega _{ad}( \sigma _{j}^{+}e^{-i\omega _{ad}^{\left( j\right) }t}+\sigma _{j}^{-}e^{i\omega _{ad}^{\left( j\right) }t})\label{qudong-a}.
     	\end{align}
     \end{subequations}
     Here, $\Omega _{cd}$ ($\Omega _{ad}$) and $\omega _{cd}$ ($\omega _{ad}^{\left( j\right)}$) are, respectively, the driving amplitude and driving frequency of the cavity driving (the driving of the $j$th atom). In particular, we consider the strong driving regime such that other physical processes can be neglected during the driving process, then the atoms can be excited on demand. For driving the cavity field into the number states, we consider the strong-driving and short-time regime, then the cavity field can be prepared into the number states through a sequence of excitation processes $\left\vert 0\right\rangle \rightarrow \left\vert 1\right\rangle \rightarrow
     \left\vert 2\right\rangle \rightarrow \dots\rightarrow \left\vert
     n\right\rangle $. With these methods, the system can be prepared into the proper basis states in the arbitrary-excitation subspaces.
     
     Below, we will study the dark-state effect and its characterization in the coupled cavity-Rydberg-atom systems. Concretely, we will consider the case of $N=2,3,4$. For keeping notation concise, some notations are commonly used in these sections. Therefore, we should point out that the notations keep consistent in each section.

	\section{Dark states in the two-atom case}\label{sec3}
	In this section, we study the dark-state effect in the system of two Rydberg atoms coupled to the cavity field, which is described by the Hamiltonian $\hat{H}_{[2]}$ [$N=2$ for Eq.~(\ref{Hn})]. Here, we consider a simplified case where the dipole-dipole interaction strength is approximated as a constant $V_{12}=V_{dd}$. Concretely, we study the dark states in the single-excitation subspace ($n=1<N=2$).
	We also investigate the characterization of the dark states in the open-system case.

	\subsection{Dark states in the single-excitation subspace}

In the single-excitation subspace, there exist three basis states $\left\{
\left\vert 1,g,g\right\rangle ,\left\vert 0,e,g\right\rangle ,\left\vert
0,g,e\right\rangle \right\} $ for the two-atom system. According to the involved cavity photon number, there is one upper state $\left\vert 1,g,g\right\rangle $ and
two lower states $\left\vert 0,e,g\right\rangle $ and $\left\vert
0,g,e\right\rangle .$ We can divide these three basis states
into two sub-components: the upper-state component $\left\vert
1,g,g\right\rangle $ and the lower-state component $\left\{ \left\vert
0,e,g\right\rangle ,\left\vert 0,g,e\right\rangle \right\} .$
We define the vectors for these basis states as $\vert u_{1}\rangle=\left\vert
1,g,g\right\rangle =\left( 1,0,0\right) ^{T}$, $\vert l_{1}\rangle=\left\vert 0,e,g\right\rangle
=\left( 0,1,0\right) ^{T}$, and $\vert l_{2}\rangle=\left\vert 0,g,e\right\rangle =\left(
0,0,1\right) ^{T}$. Then, in the single-excitation subspace, the Hamiltonian $%
\hat{H}_{[2]}$ can be expressed as%
\begin{equation}
	H_{[2]}^{\left( 1\right) }=\left(
	\begin{array}{c|c}
		\mathbf{U}_{\left[ 2\right] }^{\left( 1\right) } & \mathbf{C}_{\left[ 2%
			\right] }^{\left( 1\right) } \\ \hline
		\left( \mathbf{C}_{\left[ 2\right] }^{\left( 1\right) }\right) ^{\dag } &
		\mathbf{L}_{\left[ 2\right] }^{\left( 1\right) }%
	\end{array}%
	\right) =\left(
	\begin{array}{c|cc}
		-\Delta _{a} & g_{1} & g_{2} \\ \hline
		g_{1} & 0 & V_{dd} \\
		g_{2} & V_{dd} & 0%
	\end{array}%
	\right) ,  \label{H21}
\end{equation} 
where $\mathbf{U}_{\left[ 2\right] }^{\left( 1\right) }$ and $\mathbf{L}_{\left[ 2\right] }^{\left( 1\right) }$ are the submatrices related to the upper- and lower-state components in the single-excitation subspace, respectively, and $\mathbf{C}_{\left[ 2\right] }^{\left( 1\right) }$ is the
coupling matrix describing the couplings between these two components. It
should be noted that the superscript \textquotedblleft $\left( 1\right) $%
\textquotedblright\ and the subscript \textquotedblleft $\left[ 2\right] $%
\textquotedblright\ in Eq.~(\ref{H21}) are introduced to denote the single-excitation subspace
and two Rydberg atoms coupled to the cavity field, respectively. We consider that the coupling strengths $g_{1}$ and $g_{2}$, as well as the dipole-dipole interaction strength $V_{dd}$ are non-zero for avoiding change of the coupling structure of the system.

To analyze the dark states, we need to transform the Hamiltonian matrix in Eq.~(\ref{H21}) into an arrowhead matrix. By diagonalizing the lower-state submatrix $\mathbf{L}_{\left[ 2\right]
}^{\left( 1\right) }$ with the unitary matrix
\begin{equation}\label{sl2-1}
	\mathbf{S}_{l}=\left(
	\begin{array}{cc}
		1/{\sqrt{2}} & 1/{\sqrt{2}} \\
		-1/{\sqrt{2}} & 1/{\sqrt{2}}%
	\end{array}%
	\right),
\end{equation}
the Hamiltonian can be transformed into an arrowhead matrix%
\begin{equation}\label{h2-1arrow}
	\tilde{H}_{[2]}^{\left( 1\right) }=\left(
		\begin{array}{c|c}
			\mathbf{U}_{\left[ 2\right] }^{\left( 1\right) } & \mathbf{\tilde{C}}%
			_{\left[ 2\right] }^{\left( 1\right) } \\ \hline
			\left( \mathbf{\tilde{C}}_{\left[ 2\right] }^{\left( 1\right) }\right)
			^{\dag } & \mathbf{\tilde{L}}_{\left[ 2\right] }^{\left( 1\right) }%
		\end{array}%
		\right) =\left(
	\begin{array}{c|cc}
		-\Delta _{a} & G_{1} &G_{2} 
		\\ \hline
		 G_{1}& V_{dd} & 0 \\
		 G_{2}& 0 & -V_{dd}%
	\end{array}%
	\right) ,
\end{equation}
where the coupling strengths are introduced as $G_{1}=({g_{1}+g_{2}})/{\sqrt{2}}$ and $G_{2}=({-g_{1}+g_{2}})/{\sqrt{2}}$.
Here, these three new basis states for the Hamiltonian $\tilde{H}_{[2]}^{\left( 1\right)}$ are given by%
\begin{subequations}\label{two-state}
	\begin{align}
		\vert u_{1}\rangle=&\left\vert 1,g,g\right\rangle ,\\
		\vert L _{\left[2\right] }^{\left( 1\right) }\left(1\right)\rangle=&\frac{1}{\sqrt{2}}\left\vert 0\right\rangle\left( \left\vert e,g\right\rangle +\left\vert g,e\right\rangle \right) ,\\
		\vert L _{\left[2\right] }^{\left( 1\right) }\left(2\right)\rangle=&\frac{1}{\sqrt{2}}\left\vert 0\right\rangle\left(-\left\vert e,g\right\rangle +\left\vert g,e\right\rangle \right) .
	\end{align} 
\end{subequations}
The two states $\vert L _{\left[2\right] }^{\left( 1\right) }\left(1\right)\rangle$ and $\vert L _{\left[2\right] }^{\left( 1\right) }\left(2\right)\rangle$ are the eigenstates of the lower-state submatrix $\mathbf{\tilde{L}}_{\left[ 2\right] }^{\left( 1\right) }$, with the corresponding eigenvalues $V_{dd}$ and $-V_{dd}$.

The dark states in this case can be obtained by analyzing Eq.~(\ref{h2-1arrow}) with the arrowhead-matrix method~\cite{huang2023,zhao2026}.

(1) Consider the case of zero coupling column vector:
(i) When $g_{1}=-g_{2}$, the corresponding coupling strength $G_{1}$ between the lower state $\vert L _{\left[2\right] }^{\left( 1\right) }\left(1\right)\rangle$ and the upper state $\vert u_{1}\rangle$ is zero, then $\vert L _{\left[2\right] }^{\left( 1\right) }\left(1\right)\rangle$ becomes a dark state.
(ii) When $g_{1}=g_{2}$, we have $G_{2}=0$, then the state $\vert L _{\left[2\right] }^{\left( 1\right) }\left(2\right)\rangle$ is decoupled from the upper state $\vert u_{1}\rangle$ and becomes a dark state. We point out that these two states $\vert L _{\left[2\right] }^{\left( 1\right) }(1)\rangle$ and $\vert L _{\left[2\right] }^{\left( 1\right) }(2)\rangle$ are the Bell states~\cite{bell}, which are maximally entangled states involving two atoms.

(2) Consider the case of degenerate lower-state subspace: For avoiding change of the coupling structure of the system, we consider the case of $V_{dd}\neq 0$, then there is no degeneracy in the lower states.
	
\subsection{Characterization of the dark states}
	
In this section, we study the characterization of the dark states in the two-atom case.
In the open-system case, the dynamics of this system is govern by the Lindblad quantum master equation. As we studied in the above section, the dark-state effect only appears in the single-excitation subspace for the two-atom case. For simplicity, here we only show the energy levels of the system within the ground state and single-excitation subspaces. This is reasonable because there is no driving and the environment is a vacuum bath.

\begin{figure}[t!]
	\centering\includegraphics[width=0.48\textwidth]{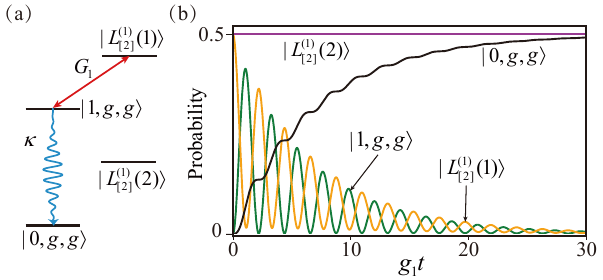}  
	\caption{(a) Energy-level diagram of the coupled cavity-two-atom system confined in the zero- and single-excitation subspaces when $g_{1}=g_{2}$. The state $\vert L _{\left[2\right] }^{\left( 1\right) }\left(2\right)\rangle $ is the dark state in this case.
	(b) The occupation probabilities of the states $\vert 1,g,g\rangle $ (green), $\vert L _{\left[2\right] }^{\left( 1\right) }\left(1\right)\rangle $ (yellow), $\vert L _{\left[2\right] }^{\left( 1\right) }\left(2\right)\rangle $ (purple), and $\vert 0,g,g\rangle $ (black) as functions of time in the open-system case. The initial state is $\vert 0,e,g\rangle $. Other parameters used are $\Delta _{a}/g_{1}=0$, $g_{2}/g_{1}=1$, $V_{dd}/g_{1}=0.5$, and $\kappa/g_{1}=0.3$.  }
	\label{two-open}
\end{figure}

Based on the Hamiltonian in Eq.~(\ref{h2-1arrow}), we plot the energy-level diagram in the zero- and single-excitation subspaces when $g_{1}=g_{2}$ in Fig.~\ref{two-open}(a). In this case, only the dressed lower state $\vert L _{\left[2\right] }^{\left( 1\right) }\left(1\right)\rangle $ is coupled to the upper state $|1,g,g\rangle$, while the other dressed lower state $\vert L _{\left[2\right] }^{\left( 1\right) }\left(2\right)\rangle $ is the dark state and here it is decoupled from the upper state $|1,g,g\rangle$. In addition, the upper state $|1,g,g\rangle$ is connected to the ground state $|0,g,g\rangle$ through the cavity-field dissipation. Since the dark state is decoupled from the cavity, it can be witness by selecting proper initial states. For example, the state $|0,e,g\rangle$ is the superposition of the bright state $\vert L _{\left[2\right] }^{\left( 1\right) }\left(1\right)\rangle $ and dark state $\vert L _{\left[2\right] }^{\left( 1\right) }\left(2\right)\rangle $, i.e., $|0,e,g\rangle=(\vert L _{\left[2\right] }^{\left( 1\right) }\left(1\right)\rangle-\vert L _{\left[2\right] }^{\left( 1\right) }\left(2\right)\rangle)/\sqrt{2}$. Under the dissipation of the cavity, the bright state $\vert L _{\left[2\right] }^{\left( 1\right) }\left(1\right)\rangle $ coupled to the upper state $|1,g,g\rangle$ will be finally dissipated into the ground state $|0,g,g\rangle$; while the dark state $\vert L _{\left[2\right] }^{\left( 1\right) }\left(2\right)\rangle $ will remain unchanged. Therefore, the steady-state population of the dark state $\vert L _{\left[2\right] }^{\left( 1\right) }\left(2\right)\rangle $ will be a signature for the existence of the dark state, and this feature can be used to characterize the dark-state effect. In particular, the dark-state population can be distinguished from the ground state $|0,g,g\rangle$ by detecting the excited-state probability of the two atoms. Hence, the present dark-state characterization can be realized in experiments.

To exhibit the dark-state characterization, we plot the populations of the states $|1,g,g\rangle$, $\vert L _{\left[2\right] }^{\left( 1\right) }\left(1\right)\rangle $, $\vert L _{\left[2\right] }^{\left( 1\right) }\left(2\right)\rangle $, and $|0,g,g\rangle$ as functions of the scaled time $g_1t$ in Fig.~\ref{two-open}(b). It can be seen that the dressed lower states $\vert L _{\left[2\right] }^{\left( 1\right) }\left(1\right)\rangle $ and $\vert L _{\left[2\right] }^{\left( 1\right) }\left(2\right)\rangle $ each possess population of $1/2$ at the initial moment. As time evolves, the populations of the states $\vert 1,g,g\rangle $ and the dressed lower state $\vert L _{\left[2\right] }^{\left( 1\right) }\left(1\right)\rangle $ exhibit coherent oscillations, then decay to the ground state $\vert 0,g,g\rangle $, while the population of the dark state $\vert L _{\left[2\right] }^{\left( 1\right) }\left(2\right)\rangle $ remains unchanged all the time. Therefore, we can observe the dark state $\vert L _{\left[2\right] }^{\left( 1\right) }\left(2\right)\rangle $ by measuring the atomic populations, and there are populations in these two atoms. We should point out that the initial state $|0,e,g\rangle$ can be prepared by only driving the first atom with the Hamiltonian in Eq.~(\ref{qudong-a}). In particular, by driving the second atom and setting the initial state to $|0,g,e\rangle$, the dark-state effect can be characterized in a similar way, because the state $|0,g,e\rangle$ is also a superposition of the states $\vert L _{\left[2\right] }^{\left( 1\right) }\left(1\right)\rangle $ and $\vert L _{\left[2\right] }^{\left( 1\right) }\left(2\right)\rangle $.

	\section{Dark states in the three-atom case}\label{sec4}
	
	We now turn to the three-atom case in which the system is described by the Hamiltonian $\hat{H}_{[3]}$ [$N=3$ for Eq.~(\ref{Hn})]. Similarly, we consider a simplified case where the dipole-dipole interaction strengths are identical, namely $V_{jj'}=V_{dd}$ for $j,j'=1,2,3$ and $j\neq j'$.
	We study the dark states in both the single- and double-excitation subspaces, as well as the characterization of the corresponding dark states in the open-system case.
	
	\subsection{Dark states in both the single- and double-excitation subspaces}
	
	\subsubsection{Single-excitation subspace}
	
	In the single-excitation subspace, there are four basis states $\left\{
	\left\vert 1,g,g,g\right\rangle ,\left\vert 0,e,g,g\right\rangle ,\left\vert
	0,g,e,g\right\rangle ,\left\vert 0,g,g,e\right\rangle \right\} $ for the three-atom system. According to the involved cavity photon number, there is one upper state $%
	\vert u_{1}\rangle=\left\vert 1,g,g,g\right\rangle $ and three lower states $\vert l_{1}\rangle=\left\vert
	0,e,g,g\right\rangle$, $\vert l_{2}\rangle=\left\vert 0,g,e,g\right\rangle $, and $\vert l_{3}\rangle=\left\vert
	0,g,g,e\right\rangle .$ We can divide these four basis states
	into two sub-components: the upper-state component $\left\vert
	1,g,g,g\right\rangle $ and the lower-state component $\left\{ \left\vert
	0,e,g,g\right\rangle ,\left\vert 0,g,e,g\right\rangle ,\left\vert
	0,g,g,e\right\rangle \right\} .$
	We define the basis vectors as $\left\vert
	1,g,g,g\right\rangle =\left( 1,0,0,0\right) ^{T}$, $\left\vert
	0,e,g,g\right\rangle =\left( 0,1,0,0\right) ^{T}$, $\left\vert
	0,g,e,g\right\rangle =\left( 0,0,1,0\right) ^{T}$, and $\left\vert
	0,g,g,e\right\rangle =\left( 0,0,0,1\right) ^{T}$. Then, in the
	single-excitation subspace, the Hamiltonian $\hat{H}_{[3]}$ is expressed as%
	\begin{equation}\label{h3-1}
		H_{[3]}^{\left( 1\right) }=\left(
		\begin{array}{c|ccc}
			-\frac{3}{2}\Delta _{a} & g_{1} & g_{2} & g_{3} \\ \hline
			g_{1} & -\frac{1}{2}\Delta _{a} & V_{dd} & V_{dd} \\
			g_{2} & V_{dd} & -\frac{1}{2}\Delta _{a} & V_{dd} \\
			g_{3} & V_{dd} & V_{dd} & -\frac{1}{2}\Delta _{a}%
		\end{array}%
		\right) ,
	\end{equation}
	where we consider non-zero $g_{j}$ (for $j=1,2,3$) and $V_{dd}$ to ensure the coupling structure unchanged for the three-atom case.
	By diagonalizing the lower-state submatrix with the unitary matrix
	\begin{equation}\label{matrix-Sl-3}
		\mathbf{S}_{l}=\left(
		\begin{array}{ccc}
			1/\sqrt{3} & 1/\sqrt{3} & 1/\sqrt{3} \\
			-1/\sqrt{2} & 1/\sqrt{2} & 0 \\
			-1/\sqrt{6} & -1/\sqrt{6} & 2/\sqrt{6}%
		\end{array}%
		\right),
	\end{equation}
	the Hamiltonian $H_{[3]}^{\left( 1\right) }$ is transformed into an arrowhead matrix%
	\begin{eqnarray}\label{3-1arrow}
		\tilde{H}_{[3]}^{\left( 1\right) } &=&\left(
		\begin{array}{c|ccc}
			-\frac{3\Delta _{a}}{2} & G_{1} & G_{2} & G_{3} \\  \hline
			G_{1} & \frac{-\Delta _{a}+4V_{dd}}{2} & 0 & 0 \\
			G_{2} & 0 & -\frac{\Delta _{a}+2V_{dd}}{2} & 0 \\
			G_{3} & 0 & 0 & -\frac{\Delta _{a}+2V_{dd}}{2}%
		\end{array}%
		\right) ,
	\end{eqnarray}%
	where the coupling strengths are introduced as $G_{1}=({g_{1}+g_{2}+g_{3}})/{\sqrt{3}}$, $G_{2}=({-g_{1}+g_{2}})/{%
		\sqrt{2}}$, and $G_{3}=-({g_{1}+g_{2}-2g_{3}})/{\sqrt{6}}$. We point out that the matrix $\tilde{H}_{[3]}^{\left( 1\right)}$ is expressed with the following basis states%
	\begin{subequations}\label{three-state}
		\begin{align}
			\vert u_{1}\rangle&= \left\vert 1,g,g,g\right\rangle, \\
			\vert L _{\left[3\right] }^{\left( 1\right) }\left(1\right)\rangle &= \frac{1}{\sqrt{3}}\left\vert 0\right\rangle \left( \left\vert e,g,g\right\rangle +\left\vert g,e,g\right\rangle +\left\vert g,g,e\right\rangle \right), \\
			\vert L _{\left[3\right] }^{\left( 1\right) }\left(2\right)\rangle &= \frac{1}{\sqrt{2}}\left\vert 0\right\rangle \left( -\left\vert e,g\right\rangle +\left\vert g,e\right\rangle \right)\left\vert g\right\rangle, \\
			\vert L _{\left[3\right] }^{\left( 1\right) }\left(3\right)\rangle &= \frac{1}{\sqrt{6}}\left\vert 0\right\rangle \left( -\left\vert e,g,g\right\rangle -\left\vert g,e,g\right\rangle +2\left\vert g,g,e\right\rangle \right).
		\end{align}	
	\end{subequations}
	
	The dark states in this case can be obtained by analyzing Eq.~(\ref{3-1arrow}) with the arrowhead-matrix method.
	
	(1) Consider the case of zero coupling column vector:
	(i) When $g_1+g_2+g_3=0$, we have $G_{1}=0$, then the state $\vert L _{\left[3\right] }^{\left( 1\right) }\left(1\right)\rangle$ is decoupled from the upper state $\vert u_{1}\rangle$ and becomes a dark state. This state $\vert L _{\left[3\right] }^{\left( 1\right) }\left(1\right)\rangle$ is a $W$ state~\cite{W-GHZ} involving three atoms.
	(ii) When $g_1=g_2$, the coupling strength $G_{2}=0$, then the state $\vert L _{\left[3\right] }^{\left( 1\right) }\left(2\right)\rangle$ becomes a dark state. In this case, the state $\vert L _{\left[3\right] }^{\left( 1\right) }\left(2\right)\rangle$ is a Bell state of the first and second atoms, and the third atom is decoupled from other subsystems.
	(iii) When $g_1+g_2=2g_3$, we get $G_{3}=0$, then the state $\vert L _{\left[3\right] }^{\left( 1\right) }\left(3\right)\rangle$ becomes a dark state, which is also an entangled state involving these three atoms.
	
	(2) Consider the case of degenerate lower-state subspace:
	It can be seen from Eq.~(\ref{3-1arrow}) that the second and third eigenvalues are identical, then there is a two-dimensional degenerate lower-state subspace $\{\vert L _{\left[3\right] }^{\left( 1\right) }\left(2\right)\rangle,\vert L _{\left[3\right] }^{\left( 1\right) }\left(3\right)\rangle \}$. As a result, there exists one dark state
    \begin{align}
    	\vert D_{[3]}^{(1)}\rangle
    	&= \frac{1}{\mathcal{N}_{[3]}^{(1)}}
    	\big( G_2\vert L_{[3]}^{(1)}(3)\rangle
    	- G_3\vert L_{[3]}^{(1)}(2)\rangle \big) \notag \\
    	&= \frac{1}{\mathcal{N}_{[3]}^{(1)}}
    	\vert 0\rangle \Bigg[
    	\left(-\frac{G_2}{\sqrt{6}}+\frac{G_3}{\sqrt{2}}\right)\vert e,g,g\rangle \notag \\
    	&\quad -\left(\frac{G_2}{\sqrt{6}}+\frac{G_3}{\sqrt{2}}\right)\vert g,e,g\rangle
    	+ \frac{G_2}{\sqrt{6}}\vert g,g,e\rangle \Bigg],
    	\label{3-1dark}
    \end{align}
	where the constant $\mathcal{N}_{[3]}^{\left(
		1\right) }=(G_{2}^{2}+G_{3}^{2}) ^{1/2}$ is introduced. The dark state $\vert D_{\left[ 3\right] }^{\left( 1\right) }\rangle$ is a $W$ state~\cite{W-GHZ}, which is an entangled state involving three atoms. It should be pointed out that, when $G_{2}=0$ or $G_{3}=0$, the state $\vert D_{\left[ 3\right] }^{\left( 1\right) }\rangle$ will be reduced to $\vert L _{\left[3\right] }^{\left( 1\right) }\left(2\right)\rangle$ or $\vert L _{\left[3\right] }^{\left( 1\right) }\left(3\right)\rangle$, respectively. 
		
		Based on the above discussions, we know that, when $G_{1}=0$, there are two dark states $\vert L _{\left[3\right] }^{\left( 1\right) }\left(1\right)\rangle$ and $\vert D_{\left[ 3\right] }^{\left( 1\right) }\rangle$ in the single-excitation subspace; when $G_{1}\neq0$, then there is one dark state $\vert D_{\left[ 3\right] }^{\left( 1\right) }\rangle$.
		
	\subsubsection{Double-excitation subspace}
	
	In the double-excitation subspace, there are seven basis states $\{
	\left\vert 2,g,g,g\right\rangle ,$ $\left\vert 1,e,g,g\right\rangle ,$ $\left\vert1,g,e,g\right\rangle ,$ $\left\vert 1,g,g,e\right\rangle ,$ $\left\vert0,e,e,g\right\rangle ,$ $\left\vert 0,e,g,e\right\rangle ,$ $\left\vert0,g,e,e\right\rangle \}$. According to the involved cavity photon number, these seven states can be divided into the upper- and lower-state components. Concretely, there are four upper states $\{\vert u_{1}\rangle=\left\vert 2,g,g,g\right\rangle,$ $\vert u_{2}\rangle=\left\vert 1,e,g,g\right\rangle ,$ $\vert u_{3}\rangle=\left\vert 1,g,e,g\right\rangle,$ $\vert u_{4}\rangle=\left\vert 1,g,g,e\right\rangle \}$ and three lower states $\{\vert l_{1}\rangle=\left\vert0,e,e,g\right\rangle ,$ $\vert l_{2}\rangle=\left\vert 0,e,g,e\right\rangle ,$ $\vert l_{3}\rangle=\left\vert0,g,e,e\right\rangle \}$. By defining these basis vectors: $\left\vert 2,g,g,g\right\rangle =\left(
	1,0,0,0,0,0,0\right) ^{T}$, $\left\vert 1,e,g,g\right\rangle =\left(
	0,1,0,0,0,0,0\right) ^{T}$, $\left\vert 1,g,e,g\right\rangle =\left(
	0,0,1,0,0,0,0\right) ^{T}$, $\left\vert 1,g,g,e\right\rangle =\left(
	0,0,0,1,0,0,0\right) ^{T}$,	$\left\vert 0,e,e,g\right\rangle =\left( 0,0,0,0,1,0,0\right) ^{T}$, $\left\vert 0,e,g,e\right\rangle =\left( 0,0,0,0,0,1,0\right) ^{T}$, $\left\vert 0,g,e,e\right\rangle =\left( 0,0,0,0,0,0,1\right) ^{T}$. The Hamiltonian $\hat{H}_{[3]}$ restricted within the double-excitation subspace can be expressed as the following matrix
	\begin{equation}\label{h3-2}
		\begin{aligned}
			H_{[3]}^{(2)}
			&= \left(
			\begin{array}{c|c}
				\mathbf{U}_{[3]}^{(2)} & \mathbf{C}_{[3]}^{(2)} \\ \hline
				\left(\mathbf{C}_{[3]}^{(2)}\right)^\dag & \mathbf{L}_{[3]}^{(2)}
			\end{array}
			\right) \\
			&= \left(
			\begin{array}{cccc|ccc}
				-\dfrac{3\Delta_a}{2} & \sqrt{2}g_1 & \sqrt{2}g_2 & \sqrt{2}g_3 & 0 & 0 & 0 \\
				\sqrt{2}g_1 & -\dfrac{\Delta_a}{2} & V_{dd} & V_{dd} & g_2 & g_3 & 0 \\
				\sqrt{2}g_2 & V_{dd} & -\dfrac{\Delta_a}{2} & V_{dd} & g_1 & 0 & g_3 \\
				\sqrt{2}g_3 & V_{dd} & V_{dd} & -\dfrac{\Delta_a}{2} & 0 & g_1 & g_2 \\ \hline
				0 & g_2 & g_1 & 0 & \dfrac{\Delta_a}{2} & V_{dd} & V_{dd}\\
				0 & g_3 & 0 & g_1 & V_{dd} & \dfrac{\Delta_a}{2} & V_{dd}\\
				0 & 0 & g_3 & g_2 & V_{dd} & V_{dd} & \dfrac{\Delta_a}{2} \\
			\end{array}
			\right).
		\end{aligned}
	\end{equation}
	Similarly, here we assume that both the variable $g_{j}$ (for $j=1,2,3$) and $V_{dd}$ are non-zero for avoiding the change of the coupling configuration for the system.
	
	To analyze the dark-state effect in this system, we diagonalize the lower-state submatrix $\mathbf{L}_{[3]}^{(2)}$. By comparing the lower-state submatrix in Eq.~(\ref{h3-2}) and Eq.~(\ref{h3-1}), we find that the lower-state submatrix in Eq.~(\ref{h3-2}) can be diagonalized with the same unitary matrix given in Eq.~(\ref{matrix-Sl-3}). The diagonalized lower-state submatrix $\mathbf{\tilde{L}}_{\left[ 3\right] }^{\left( 2\right)}$ and the corresponding coupling submatrix $\mathbf{\tilde{C}}%
	_{\left[ 3\right] }^{\left( 2\right)}$ are given by%
	\begin{subequations}\label{L3-2}
		\begin{align}
			\mathbf{\tilde{L}}_{[3]}^{(2)} &= \operatorname{diag}\left( 
			\frac{\Delta_a}{2}+2V_{dd}, 
			\frac{\Delta_a}{2}-V_{dd}, 
			\frac{\Delta_a}{2}-V_{dd}
			\right), \\
			\mathbf{\tilde{C}}_{[3]}^{(2)} &= (\mathbf{G}_1, \mathbf{G}_2, \mathbf{G}_3) = \begin{pmatrix}
				0 & 0 & 0 \\
				\frac{ g_2+g_3}{\sqrt{3}} & \frac{-g_2+g_3}{\sqrt{2}} & \frac{-g_2-g_3}{\sqrt{6}} \\
				\frac{g_1+g_3}{\sqrt{3}} & \frac{-g_1}{\sqrt{2}} & \frac{-g_1+2g_3}{\sqrt{6}} \\
				\frac{g_1+g_2 }{\sqrt{3}} & \frac{g_1}{\sqrt{2}} & \frac{-g_1+2g_2}{\sqrt{6}}
			\end{pmatrix}.
		\end{align}
	\end{subequations}
	 Note that the dressed lower states of the submatrix $\mathbf{\tilde{L}}_{[3]}^{\left( 2\right)}$ are given by
	\begin{subequations}
		\begin{align}
			\vert L _{\left[3\right] }^{\left( 2\right) }\left(1\right)\rangle &= \frac{1}{\sqrt{3}}\left\vert 0\right\rangle \left( \left\vert e,e,g\right\rangle + \left\vert e,g,e\right\rangle + \left\vert g,e,e\right\rangle \right) , \\
			\vert L _{\left[3\right] }^{\left( 2\right) }\left(2\right)\rangle &= \frac{1}{\sqrt{2}}\left\vert 0\right\rangle \left\vert e\right\rangle\left( -\left\vert e,g\right\rangle + \left\vert g,e\right\rangle \right), \\
			\vert L _{\left[3\right] }^{\left( 2\right) }\left(3\right)\rangle &= \frac{1}{\sqrt{6}}\left\vert 0\right\rangle \left( -\left\vert e,e,g\right\rangle -\left\vert e,g,e\right\rangle +2 \left\vert g,e,e\right\rangle \right).
		\end{align}
	\end{subequations}

	Based on Eqs.~(\ref{L3-2}), we can analyze the dark states with the arrowhead-matrix method.
	
	(1) Consider the case of zero coupling column vector:
	We find that there are no proper non-zero parameters $g_1$, $g_2$, and $g_3$ satisfying $\mathbf{G}_i=\mathbf{0}$ for $i=1,2,3$. As a result, no dark states corresponding to the zero coupling column vectors exist in this case.
	
	(2) Consider the case of degenerate lower-state subspace:
	Though there is a two-dimensional degenerate subspace $\{\vert L _{\left[3\right] }^{\left( 2\right) }\left(2\right)\rangle ,\vert L _{\left[3\right] }^{\left( 2\right) }\left(3\right)\rangle \}$, the corresponding coupling submatrix related to the degenerate subspace is full rank for non-zero $g_j$ (for $j=1,2,3$). Therefore, there is no dark state.
	
	Based on the above analyses, we know that there is no dark state in the double-excitation subspace for the three-atom case.
	
	\begin{figure}[t!]
		\centering\includegraphics[width=0.48\textwidth]{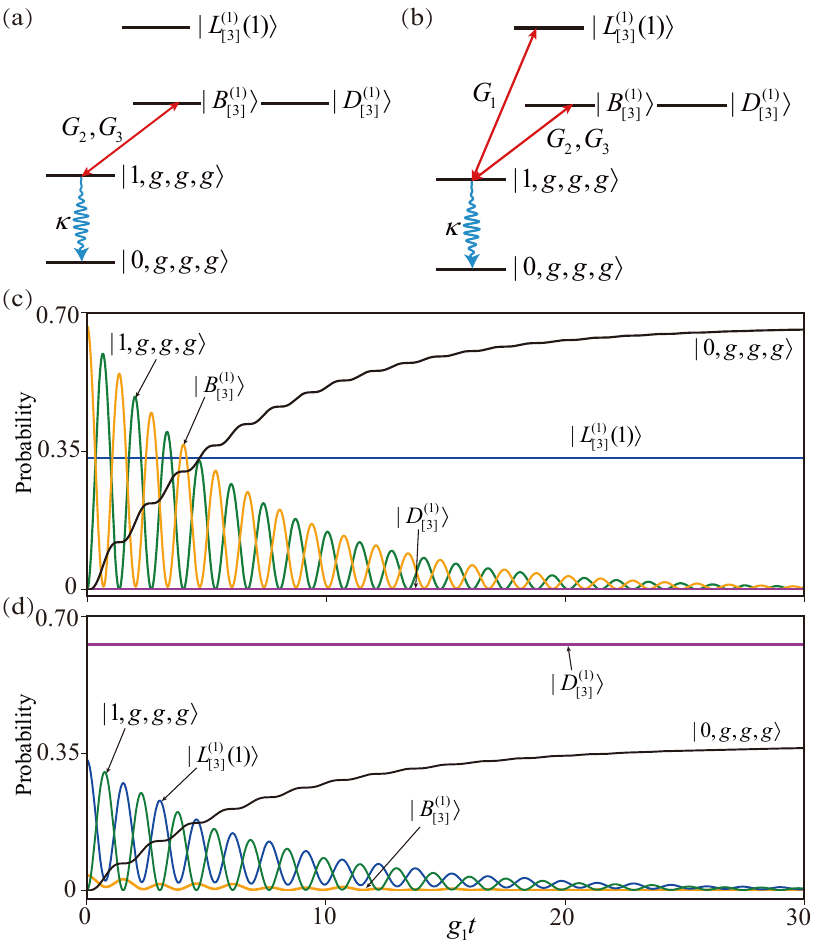}  
		\caption{Energy-level diagram of the coupled cavity-three-atom system confined in the zero- and single-excitation subspaces, when there exist (a) two dark states, (b) one dark state. (c),(d) The occupation probabilities of these states $\vert 1,g,g,g \rangle$ (green), $\vert L _{\left[3\right] }^{\left( 1\right) }\left(1\right)\rangle $ (blue), $\vert B _{\left[3\right] }^{\left( 1\right) }\rangle $ (yellow), $\vert D _{\left[3\right] }^{\left( 1\right) }\rangle $ (purple), and $\vert 0,g,g,g \rangle$ (black) as functions of time in the open-system case. The initial states are: (c) $\vert 0,g,g,e \rangle$, (d) $\vert 0,e,g,g \rangle$. The used parameters are (c) $g_{2}/g_{1}=0.9$ and $g_{3}/g_{1}=-1.9$, (d) $g_{2}/g_{1}=0.8$ and $g_{3}/g_{1}=1.5$. Other parameters used are $V_{dd}/g_{1}=0.5$, $\Delta _{a}/g_{1}=0$, and $\kappa/g_{1}=0.3$. }
		\label{three-open}
	\end{figure}
	
	\subsection{ Characterization of the dark states}
In this section, we study the characterization of dark states in the single-excitation subspace for the three-atom case.  
In the open-system case, we study the dynamics of the system based on the quantum master equation.
As we studied in the above section, there are two different situations for the dark-state effect in the single-excitation subspace for the three-atom case: (1) $G_1=0$ and (2) $G_1\neq0$. 

For case (1) $G_1=0$, we plot the energy-level diagram in the zero- and single-excitation subspaces in Fig.~\ref{three-open}(a). There are two dark states $\vert L _{\left[3\right] }^{\left( 1\right) }\left(1\right)\rangle$ and $\vert D _{\left[3\right] }^{\left( 1\right) }\rangle$, and only the bright state $\vert B _{\left[3\right] }^{\left( 1\right) }\rangle  =( G_{2}\vert L _{\left[3\right] }^{\left( 1\right) }\left(2\right)\rangle + G_{3}\vert L _{\left[3\right] }^{\left( 1\right) }\left(3\right)\rangle)/{\mathcal{N}_{\left[ 3\right] }^{\left( 1\right) }}$ is coupled to the upper state $\vert 1,g,g,g \rangle$. The upper state $\vert 1,g,g,g \rangle$ is connected to the ground state $\vert 0,g,g,g \rangle$ through the cavity-field dissipation. Owing to the decoupling of the dark state from the cavity, we can identify the dark state. We find that when the initial state is $\vert 0,g,g,e \rangle$ and $g_{1}\approx g_{2}$, the dark state $\vert L _{\left[3\right] }^{\left( 1\right) }\left(1\right)\rangle$ can be distinguished from the dark state $\vert D _{\left[3\right] }^{\left( 1\right) }\rangle$. This is because the state $\vert 0,g,g,e \rangle$ and the dark state $\vert D _{\left[3\right] }^{\left( 1\right) }\rangle$ are almost orthogonal in this case. Therefore, the presence of the dark state $\vert L _{\left[3\right] }^{\left( 1\right) }\left(1\right)\rangle$ is signaled by its steady-state population, which provides a useful means for characterizing the dark-state effect. Notably, one can differentiate the dark-state population from that of the ground state $|0,g,g,g\rangle$ via detection of the excited-state probability of the three atoms. As a result, the proposed scheme for dark-state characterization is realizable in experiments.

For case (2) $G_1\neq0$, there is one dark state $\vert D _{\left[3\right] }^{\left( 1\right) }\rangle$, as shown in Fig.~\ref{three-open}(b). Both the dressed lower state $\vert L _{\left[3\right] }^{\left( 1\right) }\left(1\right)\rangle$ and the bright state $\vert B _{\left[3\right] }^{\left( 1\right) }\rangle$ are coupled to the upper state $\vert 1,g,g,g \rangle$ which is coupled to the ground state $\vert 0,g,g,g \rangle$ via cavity-field dissipation, while the dark state $\vert D _{\left[3\right] }^{\left( 1\right) }\rangle$ is decoupled from the upper state $\vert 1,g,g,g \rangle$. Similarly, the dark-state effect can be identified from the population of the system. When the initial state is $\vert 0,e,g,g \rangle$, which is the superposition of these states $\vert L _{\left[3\right] }^{\left( 1\right) }\left(1\right)\rangle$, $\vert B _{\left[3\right] }^{\left( 1\right) }\rangle$, and $\vert D _{\left[3\right] }^{\left( 1\right) }\rangle$, the steady state of the system will be the superposition of the dark state $\vert D _{\left[3\right] }^{\left( 1\right) }\rangle$ and the ground state $\vert 0,g,g,g \rangle$. Except the dark state, all states eventually decay to the ground state $\vert 0,g,g,g \rangle$ because of the dissipation of the cavity. Hence, the existence of the dark state $\vert D _{[3]}^{(1)}\rangle$ can be revealed through its finite steady-state population, which provides a practical signature for the dark-state effect. Specifically, the population of the dark state and that of the ground state $|0,g,g,g\rangle$ can be distinguished by detecting the excited-state probability of these three atoms. This demonstrates that our dark-state characterization scheme is realizable in experiments.
 
To exhibit the dark-state characterization in case (1), we plot the populations of these states $\vert 1,g,g,g \rangle$, $\vert L _{\left[3\right] }^{\left( 1\right) }\left(1\right)\rangle $, $\vert B _{\left[3\right] }^{\left( 1\right) }\rangle$, $\vert D _{\left[3\right] }^{\left( 1\right) }\rangle $, and $\vert 0,g,g,g \rangle$ as functions of the scaled time $g_1t$ when the initial state is $\vert 0,g,g,e \rangle$ in Fig.~\ref{three-open}(c). At the initial time, populations are present for these three states $\vert L _{\left[3\right] }^{\left( 1\right) }\left(1\right)\rangle$ and $\vert B _{\left[3\right] }^{\left( 1\right) }\rangle$ while the population of the dark state $\vert D _{\left[3\right] }^{\left( 1\right) }\rangle$ is approach zero. As time evolves, we can find that the populations of states $\vert 1,g,g,g \rangle$ and $\vert B _{\left[3\right] }^{\left( 1\right) }\rangle$ exhibit oscillations, then decay to the ground state $\vert 0,g,g,g \rangle$, while the population of the dark state $\vert L _{\left[3\right] }^{\left( 1\right) }\left(1\right)\rangle$ remains unchanged all the time. Therefore, the system eventually relaxes to a steady state with population only in the dark state $\vert L _{\left[3\right] }^{\left( 1\right) }\left(1\right)\rangle$ and the ground state $\vert 0,g,g,g \rangle$, and these two states can be distinguished by measuring the atomic populations. In addition, the initial state $\vert 0,g,g,e \rangle$ can be prepared by only driving the third atom with the Hamiltonian in Eq.~(\ref{qudong-a}). All these features increase the probability for experimental implementation of this system.
 
For case (2), in Fig.~\ref{three-open}(d) we plot the populations of these states $\vert 1,g,g,g \rangle$, $\vert L _{\left[3\right] }^{\left( 1\right) }\left(1\right)\rangle $, $\vert B _{\left[3\right] }^{\left( 1\right) }\rangle$, $\vert D _{\left[3\right] }^{\left( 1\right) }\rangle $, and $\vert 0,g,g,g \rangle$ as functions of the scaled time $g_1t$ when the initial state is $\vert 0,e,g,g \rangle$. Similarly, at the initial time, populations are present for these three states  $\vert L _{\left[3\right] }^{\left( 1\right) }\left(1\right)\rangle$, $\vert B _{\left[3\right] }^{\left( 1\right) }\rangle$, and $\vert D _{\left[3\right] }^{\left( 1\right) }\rangle$. As time evolves, the populations of these states $\vert 1,g,g,g \rangle$, $\vert L _{\left[3\right] }^{\left( 1\right) }\left(1\right)\rangle$, and $ \vert B _{\left[3\right] }^{\left( 1\right) }\rangle $ exhibit oscillations, and then decay to the ground state $\vert 0,g,g,g \rangle$, while the population of the dark state $\vert D _{\left[3\right] }^{\left( 1\right) }\rangle$ stays the same throughout the dynamics. Therefore, the system eventually relaxes to a steady state with population only in the dark state $\vert D _{\left[3\right] }^{\left( 1\right) }\rangle$ and the ground state $\vert 0,g,g,g \rangle$. The dark state $\vert D _{\left[3\right] }^{\left( 1\right) }\rangle$ can be identified by measuring the atomic populations, and there exist excited-state populations in these three atoms. The initial state $\vert 0,e,g,g \rangle$ can also be prepared by only driving the first atom.

	\section{Dark states in the four-atom case}\label{sec5}
	
	In this section, we study the dark-state effect with constant dipole-dipole interaction strengths, i.e., $V_{jj^{\prime }}=V_{dd}$ for $j,j'=1$-$4$ and $j\neq j'$ in the four-atom system, which is described by the Hamiltonian $\hat{H}_{[4]}$ [$N=4$ for Eq.~(\ref{Hn})]. We study the dark states in the single-, double-, and three-excitation subspaces. We also present the characterization of the dark states in the open-system case.
	
	\subsection{Dark states in the single-, double-, and three-excitation subspaces}
	
	\subsubsection{Single-excitation subspace}
	
	In the single-excitation subspace, the basis states are given by $\{ \left\vert
	1,g,g,g,g\right\rangle ,$ $\left\vert 0,e,g,g,g\right\rangle ,$ $\left\vert
	0,g,e,g,g\right\rangle ,$ $\left\vert 0,g,g,e,g\right\rangle ,$ $\left\vert
	0,g,g,g,e\right\rangle \} $ for the four-atom system, and there is one upper state $\vert u_{1}\rangle=\left\vert1,g,g,g,g\right\rangle $ and four lower states $\{ \vert l_{1}\rangle=\left\vert0,e,g,g,g\right\rangle ,$ $\vert l_{2}\rangle=\left\vert 0,g,e,g,g\right\rangle ,$ $\vert l_{3}\rangle=\left\vert0,g,g,e,g\right\rangle ,$ $\vert l_{4}\rangle=\left\vert 0,g,g,g,e\right\rangle \} .$ We define the basis vectors: $\left\vert 1,g,g,g,g\right\rangle =\left(1,0,0,0,0\right) ^{T}$, $\left\vert 0,e,g,g,g\right\rangle =\left(0,1,0,0,0\right) ^{T}$, $\left\vert 0,g,e,g,g\right\rangle =\left(0,0,1,0,0\right) ^{T}$, $\left\vert 0,g,g,e,g\right\rangle =\left(0,0,0,1,0\right) ^{T}$, $\left\vert 0,g,g,g,e\right\rangle =\left(0,0,0,0,1\right) ^{T}.$ Then the Hamiltonian $\hat{H}_{[4]}$ in the single-excitation subspace can be expressed as%
	\begin{equation}\label{h4-1}
		H_{[4]}^{\left( 1\right) }=\left(
		\begin{array}{c|cccc}
			-2\Delta _{a} & g_{1} & g_{2} & g_{3} & g_{4} \\  \hline
			g_{1} & -\Delta _{a} & V_{dd} & V_{dd} & V_{dd} \\
			g_{2} & V_{dd} & -\Delta _{a} & V_{dd} & V_{dd} \\
			g_{3} & V_{dd} & V_{dd} & -\Delta _{a} & V_{dd} \\
			g_{4} & V_{dd} & V_{dd} & V_{dd} & -\Delta _{a}%
		\end{array}%
		\right) .
	\end{equation}
	Similarly, we consider that both the coupling strength $g_{j}$ (for $j=1$-$4$) and $V_{dd}$ are non-zero to prevent change of the coupling configuration for the system.
	
	By diagonalizing the lower-state submatrix with the unitary matrix
	\begin{equation}\label{matrix-sl-4}
		\mathbf{S}_{l}=\left(
		\begin{array}{cccc}
			1/2 & 1/2 & 1/2 & 1/2 \\
			-1/\sqrt{2} & 1/\sqrt{2} & 0 & 0 \\
			-1/\sqrt{6} & -1/\sqrt{6} & 2/\sqrt{6} & 0 \\
			-\sqrt{3}/6 & -\sqrt{3}/6 & -\sqrt{3}/6 &\sqrt{3}/2 %
		\end{array}%
		\right),
	\end{equation}
	the Hamiltonian $H_{[4]}^{\left( 1\right) }$ becomes an arrowhead matrix
	\begin{equation}\label{h4-1arrow}
		\tilde{H}_{[4]}^{\left( 1\right) }=\left(
		\begin{array}{c|c}
			\mathbf{U}_{\left[ 4\right] }^{\left( 1\right) } & \mathbf{\tilde{C}}%
			_{\left[ 4\right] }^{\left( 1\right) } \\ \hline
			\left( \mathbf{\tilde{C}}_{\left[ 4\right] }^{\left( 1\right) }\right)
			^{\dag } & \mathbf{\tilde{L}}_{\left[ 4\right] }^{\left( 1\right) }%
		\end{array}%
		\right) ,
	\end{equation}%
	where these submatrices are given by%
	\begin{subequations}
		\begin{align}
			\mathbf{U}_{\left[ 4\right] }^{\left( 1\right) }=& -2\Delta
			_{a} ,\\
			\mathbf{\tilde{L}}_{\left[ 4\right] }^{\left( 1\right) }=&\text{diag}\left( -\Delta_{a}+3V_{dd},-\Delta _{a}-V_{dd},-\Delta _{a}-V_{dd},-\Delta
			_{a}-V_{dd}\right) ,\\
			\mathbf{\tilde{C}}_{\left[ 4\right] }^{\left( 1\right) }=&(G_{1}, G_{2}, G_{3}, G_{4}) .
		\end{align}%
	\end{subequations}
	Here, we introduce the coupling strengths $G_{1}=\left(
	g_{1}+g_{2}+g_{3}+g_{4}\right) /2$, $G_{2}=\left( -g_{1}+g_{2}\right) /\sqrt{2}$, $G_{3}=\left( -g_{1}-g_{2}+2g_{3}\right) /\sqrt{6}$, and $G_{4}=\left(
	-g_{1}-g_{2}-g_{3}+3g_{4}\right) /2\sqrt{3}$.
	We point out that the five new basis states of the Hamiltonian $\tilde{H}_{[4]}^{\left( 1\right)}$ are given by%
		\begin{subequations}\label{four-state}
		\begin{align}
			\vert u_{1}\rangle &= \vert 1,g,g,g,g\rangle, \\
			\vert L _{\left[4\right] }^{\left( 1\right) }\left(1\right)\rangle &= \frac{1}{2}\vert 0\rangle( \vert e,g,g,g\rangle + \vert g,e,g,g\rangle  + \vert g,g,e,g\rangle \notag \\
			&\quad+ \vert g,g,g,e\rangle ), \\
			\vert L _{\left[4\right] }^{\left( 1\right) }\left(2\right)\rangle &= \frac{1}{\sqrt{2}}\vert 0\rangle( -\vert e,g\rangle + \vert g,e\rangle )\vert g,g\rangle, \\
			\vert L _{\left[4\right] }^{\left( 1\right) }\left(3\right)\rangle &= \frac{1}{\sqrt{6}}\vert 0\rangle( -\vert e,g,g\rangle -\vert g,e,g\rangle  +2 \vert g,g,e\rangle )\vert g\rangle, \\
			\vert L _{\left[4\right] }^{\left( 1\right) }\left(4\right)\rangle &= \frac{\sqrt{3}}{6}\vert 0\rangle( -\vert e,g,g,g\rangle -\vert g,e,g,g\rangle - \vert g,g,e,g\rangle \notag \\
			&\quad +3\vert g,g,g,e\rangle ).
		\end{align}
	\end{subequations}
	
	The dark states in this case can be obtained by analyzing Eq.~(\ref{h4-1arrow}) with the arrowhead-matrix method.
	
	(1) Consider the case of zero coupling column vector:
	(i) When $g_1+g_2+g_3+g_4=0$, we have $G_{1}=0$, then the state $\vert L _{\left[4\right] }^{\left( 1\right) }\left(1\right)\rangle$ is decoupled from the upper state $\vert u_{1}\rangle$ and becomes a dark state. This state $\vert L _{\left[4\right] }^{\left( 1\right) }\left(1\right)\rangle$ is a $W$ state for four atoms.
	(ii) When $g_1=g_2$, the coupling strength $G_{2}=0$. In this case, the corresponding state $\vert L _{\left[4\right] }^{\left( 1\right) }\left(2\right)\rangle$ is decoupled from the upper state $\vert u_{1}\rangle$ and becomes a dark state. In this case, the dark state $\vert L _{\left[4\right] }^{\left( 1\right) }\left(2\right)\rangle$ is a Bell state for the first and second atoms, and the third and fourth atoms are decoupled from other subsystems.
	(iii) When $g_1+g_2=2g_3$, we get $G_{3}=0$, then the state $\vert L _{\left[4\right] }^{\left( 1\right) }\left(3\right)\rangle$ becomes a dark state, which is an entangled state involving the former three atoms.
	(iv) When $g_1+g_2+g_3=3g_4$, we have $G_{4}=0$, then the state $\vert L _{\left[4\right] }^{\left( 1\right) }\left(4\right)\rangle$ becomes a dark state, and it is an entangled state involving four atoms.
	
	(2) Consider the case of degenerate lower-state subspace:
	There is a three-dimensional degenerate subspace $\{\vert L _{\left[4\right] }^{\left( 1\right) }\left(2\right)\rangle ,\vert L _{\left[4\right] }^{\left( 1\right) }\left(3\right)\rangle,\vert L _{\left[4\right] }^{\left( 1\right) }\left(4\right)\rangle \}$, and according to arrowhead-matrix method, there exist two dark states
	\begin{subequations}\label{4-1dark}
		\begin{align}
			\vert D_{\left[ 4\right] }^{\left( 1\right) }\left( 1\right)
			\rangle =&\frac{1}{\sqrt{G_{2}^{2}+G_{3}^{2}}}( G_{3}\vert L _{\left[4\right] }^{\left( 1\right) }\left(2\right)\rangle -G_{2}\vert L _{\left[4\right] }^{\left( 1\right) }\left(3\right)\rangle ) ,\\
			\vert D_{\left[ 4\right] }^{\left( 1\right) }\left( 2\right)
			\rangle =&\frac{1}{\sqrt{G_{2}^{2}+G_{4}^{2}}}( G_{4}\vert L _{\left[4\right] }^{\left( 1\right) }\left(2\right)\rangle -G_{2}\vert L _{\left[4\right] }^{\left( 1\right) }\left(4\right)\rangle ) .
		\end{align}
	\end{subequations}
	 Note that the dark states are not unique because the linear dependence is not unique, and two linearly independent dark states span a two-dimensional subspace of dark states. These two dark states can be orthogonalized by using the Gram-Schmidt orthogonalization,	
	 \begin{subequations}\label{4-1dark-orth}
	 	\begin{align}
	 		\vert \tilde{D}_{[4]}^{(1)}(1)\rangle
	 		&= \frac{1}{\sqrt{G_{2}^{2}+G_{3}^{2}}} ( G_{3}\vert L _{\left[4\right] }^{\left( 1\right) }\left(2\right)\rangle - G_{2}\vert L _{\left[4\right] }^{\left( 1\right) }\left(3\right)\rangle ) \notag \\
	 		&=\frac{1}{\sqrt{G_{2}^{2}+G_{3}^{2}}}\vert 0\rangle \Bigg(\frac{\sqrt{2}G_{2}-\sqrt{6}G_{3}}{2\sqrt{3}} \vert e,g,g\rangle \notag \\
	 		&\quad+\frac{\sqrt{2}G_{2}+\sqrt{6}G_{3}}{2\sqrt{3}} \vert g,e,g\rangle-\frac{2G_{2}}{\sqrt{6}}\vert g,g,e\rangle \bigg)\vert g\rangle, \\	 				
	 		\vert \tilde{D}_{[4]}^{(1)}(2)\rangle
	 		&=\frac{1}{\sqrt{G_{2}^{2}+G_{3}^{2}}\sqrt{G_{2}^{2}+G_{3}^{2}+G_{4}^{2}}}( G_{2}G_{4}\vert L _{\left[4\right] }^{\left( 1\right) }\left(2\right)\rangle \notag \\
	 		&\quad+ G_{3}G_{4} \vert L _{\left[4\right] }^{\left( 1\right) }\left(3\right)\rangle - ( G_{2}^{2}+G_{3}^{2}) \vert L _{\left[4\right] }^{\left( 1\right) }\left(4\right)\rangle ) \notag \\
	 		&=\frac{1}{\sqrt{G_{2}^{2}+G_{3}^{2}}\sqrt{G_{2}^{2}+G_{3}^{2}+G_{4}^{2}}}\vert 0\rangle \notag \\
	 		&\quad \times \Bigg( \frac{\sqrt{3}(G_{2}^{2}+G_{3}^{2})-3\sqrt{2}G_{2}G_{4}-\sqrt{6}G_{3}G_{4}}{6} \vert e,g,g,g\rangle \notag \\
	 		&\quad+\frac{3\sqrt{2}G_{2}G_{4}-\sqrt{6}G_{3}G_{4}+\sqrt{3}(G_{2}^{2}+G_{3}^{2})}{6}\vert g,e,g,g\rangle\notag \\
	 		&\quad+\frac{2\sqrt{6}G_{3}G_{4}+\sqrt{3}(G_{2}^{2}+G_{3}^{2})}{6} \vert g,g,e,g\rangle \notag \\
	 		&\quad -\frac{3\sqrt{3}(G_{2}^{2}+G_{3}^{2})}{6}\vert g,g,g,e\rangle \Bigg).
	 	\end{align}
	 \end{subequations}
	 			
	 These two states in Eqs.~(\ref{4-1dark-orth}) can be employed as a complete set of orthogonal dark states for this subspace, and any unitary transformation of them can express a new set of orthogonal dark states. Therefore, the orthogonal dark states are not unique, but the dark-state subspace is unique.
	 It can be seen from Eqs.~(\ref{4-1dark-orth}) that, when $G_2=0$ or $G_3=0$, the state $\vert \tilde{D}_{[4]}^{(1)}(1)\rangle$ is reduced to $\vert L _{\left[4\right] }^{\left( 1\right) }\left(2\right)\rangle$ or $\vert L _{\left[4\right] }^{\left( 1\right) }\left(3\right)\rangle$, respectively. In addition, when $G_4=0$, the state $\vert \tilde{D}_{[4]}^{(1)}(2)\rangle$ is reduced to $\vert L _{\left[4\right] }^{\left( 1\right) }\left(4\right)\rangle$. 
	 
	 Based on the above discussions, we can see that, when $G_1=0$, there exists a single dark state $\vert L _{\left[4\right] }^{\left( 1\right) }\left(1\right)\rangle$ with eigenenergy $-\Delta_{a}+3V_{dd}$, and a two-dimensional degenerate dark-state subspace with eigenenergy $-\Delta_{a}-V_{dd}$ with the orthogonal basis states $\vert \tilde{D}_{[4]}^{(1)}(1)\rangle$ and $\vert \tilde{D}_{[4]}^{(1)}(2)\rangle$. When $G_2=0$, there exists a two-dimensional degenerate dark-state subspace with eigenenergy $-\Delta_{a}-V_{dd}$, and the two dark states $\vert L _{\left[4\right] }^{\left( 1\right) }\left(2\right)\rangle$ and $\vert \tilde{D}_{[4]}^{(1)}(2)\rangle$ form the basis states of this degenerate subspace. When $G_3=0$, the two dark states $\vert L _{\left[4\right] }^{\left( 1\right) }\left(3\right)\rangle$ and $\vert \tilde{D}_{[4]}^{(1)}(2)\rangle$ form a degenerate dark-state subspace with eigenenergy $-\Delta_{a}-V_{dd}$. When $G_4=0$, the two dark states $\vert L _{\left[4\right] }^{\left( 1\right) }\left(4\right)\rangle$ and $\vert \tilde{D}_{[4]}^{(1)}(1)\rangle$ from a degenerate dark-state subspace with the eigenenergy $-\Delta_{a}-V_{dd}$. For these three cases $G_{2,3,4}=0$, there exists a two-dimensional degenerate dark-state subspace. When $G_{j=1\text{-}4}\neq0$, then there exists a degenerate dark-state subspace with the basis states $\vert \tilde{D}_{[4]}^{(1)}(1)\rangle$ and $\vert \tilde{D}_{[4]}^{(1)}(2)\rangle$.

	\subsubsection{Double-excitation subspace}
	
	In the double-excitation subspace, there are five upper states $\{\vert u_{1}\rangle=\left\vert
	2,g,g,g,g\right\rangle $, $\vert u_{2}\rangle=\left\vert 1,e,g,g,g\right\rangle $, $\vert u_{3}\rangle=\left\vert
	1,g,e,g,g\right\rangle $, $\vert u_{4}\rangle=\left\vert 1,g,g,e,g\right\rangle $, $\vert u_{5}\rangle=\left\vert 1,g,g,g,e\right\rangle \}$ and six lower states $\{\vert l_{1}\rangle=\left\vert 0,e,e,g,g\right\rangle
	$, $\vert l_{2}\rangle=\left\vert 0,e,g,e,g\right\rangle $, $\vert l_{3}\rangle=\left\vert 0,e,g,g,e\right\rangle
	$, $\vert l_{4}\rangle=\left\vert 0,g,e,e,g\right\rangle $, $\vert l_{5}\rangle=\left\vert 0,g,e,g,e\right\rangle
	$, $\vert l_{6}\rangle=\left\vert 0,g,g,e,e\right\rangle \}$. We arrange the basis states in order and define the basis vectors corresponding to these basis states as $( 1,0,0,\dots,0,0,0)^{T}, ( 0,1,0,\dots,0,0,0)^{T}$, $\ldots$, $( 0,0,0,\dots,0,1,0)^{T},$ and $( 0,0,0,\dots,0,0,1)^{T}$. The Hamiltonian $\hat{H}_{[4]}$ in the double-excitation subspace can be expressed as
	\begin{equation}
		H_{[4]}^{\left( 2\right) }=\left(
		\begin{array}{c|c}
			\mathbf{U}_{\left[4\right] }^{\left( 2\right) } & \mathbf{C}_{\left[ 4%
				\right] }^{\left( 2\right) } \\ \hline
			\left( \mathbf{C}_{\left[ 4\right] }^{\left( 2\right) }\right) ^{\dag } &
			\mathbf{L}_{\left[ 4\right] }^{\left(2\right) }%
		\end{array}%
		\right) , 
	\end{equation}
	where these submatrices are given by
	\begin{subequations}
		\begin{align}
			\mathbf{U}_{[4]}^{(2)} &= 
			\begin{pmatrix} 
				-2\Delta _{a} & \sqrt{2}g_{1} & \sqrt{2}g_{2} & \sqrt{2}g_{3} & \sqrt{2}g_{4} \\
				\sqrt{2}g_{1} & -\Delta _{a} & V_{dd} & V_{dd} & V_{dd} \\
				\sqrt{2}g_{2} & V_{dd} & -\Delta _{a} & V_{dd} & V_{dd} \\
				\sqrt{2}g_{3} & V_{dd} & V_{dd} & -\Delta _{a} & V_{dd} \\
				\sqrt{2}g_{4} & V_{dd} & V_{dd} & V_{dd} & -\Delta _{a}
			\end{pmatrix},\\
			\mathbf{L}_{[4]}^{(2)} &= 
			\begin{pmatrix} 
				0 & V_{dd} & V_{dd} & V_{dd} & V_{dd} & 0 \\
				V_{dd} & 0 & V_{dd} & V_{dd} & 0 & V_{dd} \\
				V_{dd} & V_{dd} & 0 & 0 & V_{dd} & V_{dd} \\
				V_{dd} & V_{dd} & 0 & 0 & V_{dd} & V_{dd} \\
				V_{dd} & 0 & V_{dd} & V_{dd} & 0 & V_{dd} \\
				0 & V_{dd} & V_{dd} & V_{dd} & V_{dd} & 0 
			\end{pmatrix},\\
			\mathbf{C}_{[4]}^{(2)}&= \begin{pmatrix}
				0 & 0 & 0 & 0 & 0 & 0 \\
				g_{2} & g_{3}& g_{4} & 0 & 0 & 0 \\
				g_{1} & 0 & 0 & g_{3} & g_{4} & 0\\
				0 & g_{1} & 0 & g_{2} & 0 & g_{4} \\
				0 & 0 & g_{1} & 0 & g_{2} & g_{3} 
			\end{pmatrix}.
		\end{align}
	\end{subequations}
	
	Then we diagonalize the lower-state submatrix $\mathbf{L}_{[4]}^{(2)}$ with the unitary matrix
	\begin{equation}
		\small{\mathbf{S}_{l}=\left(
		\begin{array}{cccccc}
			1/\sqrt{6} & 1/\sqrt{6} & 1/\sqrt{6} & 1/\sqrt{6} & 1/\sqrt{6} & 1/\sqrt{6}
			\\
			1/2 & 0 & -1/2 & -1/2 & 0 & 1/2 \\
			-\sqrt{3}/6 & 1/\sqrt{3} & -\sqrt{3}/6 & -\sqrt{3}/6 & 1/\sqrt{3} & -\sqrt{3}%
			/6 \\
			-1/\sqrt{2} & 0 & 0 & 0 & 0 & 1/\sqrt{2} \\
			0 & -1/\sqrt{2} & 0 & 0 & 1/\sqrt{2} & 0 \\
			0 & 0 & -1/\sqrt{2} & 1/\sqrt{2} & 0 & 0%
		\end{array}%
		\right)},
	\end{equation}
	and the Hamiltonian becomes
	\begin{equation}\label{h4-2arrow}
		\tilde{H}_{[4]}^{\left( 2\right) }=\left(
		\begin{array}{c|c}
			\mathbf{U}_{\left[ 4\right] }^{\left( 2\right) } & \mathbf{\tilde{C}}%
			_{\left[ 4\right] }^{\left( 2\right) } \\ \hline
			\left( \mathbf{\tilde{C}}_{\left[ 4\right] }^{\left( 2\right) }\right)
			^{\dag } & \mathbf{\tilde{L}}_{\left[ 4\right] }^{\left( 2\right) }%
		\end{array}%
		\right) ,
	\end{equation}%
	where these submatrices $\mathbf{\tilde{L}}_{[4]}^{(2)}$ and $\mathbf{\tilde{C}}_{[4]}^{(2)}$ are given by
		\begin{subequations}
			\begin{align}
				\mathbf{\tilde{L}}_{[4]}^{(2)} &= \operatorname{diag}\left(
				4V_{dd},-2V_{dd},-2V_{dd},0,0,0\right),\\
				\mathbf{\tilde{C}}_{[4]}^{(2)} &= (\mathbf{G}_1, \mathbf{G}_2, \mathbf{G}_3, \mathbf{G}_4, \mathbf{G}_5, \mathbf{G}_6) \notag \\
				&= \begin{pmatrix}
					0 & 0 & 0 & 0 & 0 & 0 \\
					\frac{g_{2}+g_{3}+g_{4}}{\sqrt{6}} & \frac{g_{2}-g_{4}}{2} & 
					-\frac{g_{2}-2g_{3}+g_{4}}{2\sqrt{3}} & -\frac{g_{2}}{\sqrt{2}} & -\frac{g_{3}}{\sqrt{2}} & -\frac{g_{4}}{\sqrt{2}} \\
					\frac{g_{1}+g_{3}+g_{4}}{\sqrt{6}} & \frac{g_{1}-g_{3}}{2} & 
					-\frac{g_{1}+g_{3}-2g_{4}}{2\sqrt{3}} & -\frac{g_{1}}{\sqrt{2}} & \frac{g_{4}}{\sqrt{2}} & \frac{g_{3}}{\sqrt{2}} \\
					\frac{g_{1}+g_{2}+g_{4}}{\sqrt{6}} & \frac{-g_{2}+g_{4}}{2} &
					\frac{2g_{1}-g_{2}-g_{4}}{2\sqrt{3}} & \frac{g_{4}}{\sqrt{2}} & -\frac{g_{1}}{\sqrt{2}} & \frac{g_{2}}{\sqrt{2}} \\
					\frac{g_{1}+g_{2}+g_{3}}{\sqrt{6}} & \frac{-g_{1}+g_{3}}{2} &
					-\frac{g_{1}-2g_{2}+g_{3}}{2\sqrt{3}} & \frac{g_{3}}{\sqrt{2}} & \frac{g_{2}}{\sqrt{2}} & -\frac{g_{1}}{\sqrt{2}}
				\end{pmatrix}.
			\end{align}
		\end{subequations}
	Here, we introduce dressed lower states of the Hamiltonian $\tilde{H}_{[4]}^{\left( 2\right)}$ as
	\begin{subequations}
		\begin{align}
			\vert L _{\left[4\right] }^{\left( 2\right) }\left(1\right)\rangle &= \frac{1}{\sqrt{6}}\vert 0\rangle ( \vert e,e,g,g\rangle + \vert e,g,e,g\rangle + \vert e,g,g,e\rangle  \notag \\
			&\quad+ \vert g,e,e,g\rangle + \vert g,e,g,e\rangle + \vert g,g,e,e\rangle ), \\
			\vert L _{\left[4\right] }^{\left( 2\right) }\left(2\right)\rangle &= \frac{1}{2}\vert 0\rangle ( \vert e,e,g,g\rangle - \vert e,g,g,e\rangle - \vert g,e,e,g\rangle  \notag \\
			&\quad + \vert g,g,e,e\rangle ), \\
			\vert L _{\left[4\right] }^{\left( 2\right) }\left(3\right)\rangle &= \frac{\sqrt{3}}{6}\vert 0\rangle ( -\vert e,e,g,g\rangle + 2\vert e,g,e,g\rangle - \vert e,g,g,e\rangle\notag \\
			&\quad - \vert g,e,e,g\rangle  + 2\vert g,e,g,e\rangle - \vert g,g,e,e\rangle ), \\
			\vert L _{\left[4\right] }^{\left( 2\right) }\left(4\right)\rangle &= \frac{1}{\sqrt{2}}\vert 0\rangle ( -\vert e,e,g,g\rangle + \vert g,g,e,e\rangle ), \\
			\vert L _{\left[4\right] }^{\left( 2\right) }\left(5\right)\rangle &= \frac{1}{\sqrt{2}}\vert 0\rangle ( -\vert e,g,e,g\rangle + \vert g,e,g,e\rangle ), \\
			\vert L _{\left[4\right] }^{\left( 2\right) }\left(6\right)\rangle &= \frac{1}{\sqrt{2}}\vert 0\rangle ( -\vert e,g,g,e\rangle + \vert g,e,e,g\rangle ).
		\end{align}
		\label{4-2lower-state}
	\end{subequations}
	
	The dark states in this case can be obtained by analyzing Eq.~(\ref{h4-2arrow}) with the arrowhead-matrix method. 
	
	(1) Consider the case of zero coupling column vector:
	For non-zero $g_1$, $g_2$, $g_3$, and $g_4$, there is no dark state related to the zero coupling vector. 
	
	(2) Consider the case of degenerate lower-state subspace: There are two degenerate subspaces $\{\vert L _{\left[4\right] }^{\left( 2\right) }\left(2\right)\rangle ,\vert L _{\left[4\right] }^{\left( 2\right) }\left(3\right)\rangle\}$ and $\{\vert L _{\left[4\right] }^{\left( 2\right) }\left(4\right)\rangle ,\vert L _{\left[4\right] }^{\left( 2\right) }\left(5\right)\rangle,\vert L _{\left[4\right] }^{\left( 2\right) }\left(6\right)\rangle \}$.
	
	(i) For the two-dimensional degenerate subspace $\{\vert L _{\left[4\right] }^{\left( 2\right) }\left(2\right)\rangle ,\vert L _{\left[4\right] }^{\left( 2\right) }\left(3\right)\rangle\}$, the corresponding coupling submatrix is full rank with non-zero $g_j$ (for $j=1$-$4$). As a result, there is no dark state in this two-dimensional degenerate subspace.
	
	(ii) For the three-dimensional degenerate subspace $\{\vert L _{\left[4\right] }^{\left( 2\right) }\left(4\right)\rangle ,\vert L _{\left[4\right] }^{\left( 2\right) }\left(5\right)\rangle,\vert L _{\left[4\right] }^{\left( 2\right) }\left(6\right)\rangle \}$, the number of the dark state is equal to $3-R$, where $R$ is the rank of the submatrix related to the degenerate subspace. 
	
	(a) When $g_2=g_3$ and $g_1=-g_4$, i.e., $\mathbf{G}_4=\mathbf{G}_5 $, the rank of the corresponding coupling submatrix formed by $\mathbf{G}_4$ and $\mathbf{G}_5 $ is one. Therefore, there exists a dark state composed by the states $\vert L _{\left[4\right] }^{\left( 2\right) }\left(4\right)\rangle$ and $\vert L _{\left[4\right] }^{\left( 2\right) }\left(5\right)\rangle$,
	\begin{align}\label{4-2-1dark}
		\vert D_{[4]}^{(2)}(1)\rangle
		&= \frac{1}{\sqrt{2}}( \vert L _{\left[4\right] }^{\left( 2\right) }\left(4\right)\rangle - \vert L _{\left[4\right] }^{\left( 2\right) }\left(5\right)\rangle ) \nonumber \\
		&= \frac{1}{2}\vert 0\rangle (-\vert e,e,g,g\rangle + \vert g,g,e,e\rangle  \nonumber \\
		&\quad +\vert e,g,e,g\rangle-\vert g,e,g,e\rangle ).
	\end{align}

	(b) When $g_2=g_4$ and $g_1=-g_3$, i.e., $\mathbf{G}_4=\mathbf{G}_6$, the rank of the coupling submatrix formed by $\mathbf{G}_4$ and $\mathbf{G}_6 $ is one. Therefore, there exists the following dark state
	\begin{align}\label{4-2-2dark}
		\vert D_{[4]}^{(2)}(2)\rangle
		&= \frac{1}{\sqrt{2}}( \vert L _{\left[4\right] }^{\left( 2\right) }\left(4\right)\rangle - \vert L _{\left[4\right] }^{\left( 2\right) }\left(6\right)\rangle ) \nonumber \\
		&= \frac{1}{2}\vert 0\rangle (-\vert e,e,g,g\rangle+ \vert g,g,e,e\rangle  \nonumber \\
		&\quad +\vert e,g,g,e\rangle -\vert g,e,e,g\rangle).
	\end{align}
	
	(c) When $g_3=g_4$ and $g_1=-g_2$, i.e., $\mathbf{G}_5=\mathbf{G}_6$, the rank of the coupling submatrix formed by $\mathbf{G}_5$ and $\mathbf{G}_6 $ is one. In this case, the single dark state reads
	\begin{align}\label{4-2-3dark}
		\vert D_{[4]}^{(2)}(3)\rangle
		&= \frac{1}{\sqrt{2}}( \vert L _{\left[4\right] }^{\left( 2\right) }\left(5\right)\rangle- \vert L _{\left[4\right] }^{\left( 2\right) }\left(6\right)\rangle ) \nonumber \\
		&=\frac{1}{2}\vert 0\rangle (-\vert e,g,e,g\rangle+\vert g,e,g,e\rangle  \nonumber \\
		&\quad +\vert e,g,g,e\rangle -\vert g,e,e,g\rangle).
	\end{align}
	
	(d) When $-g_1=g_2=g_3=g_4$, i.e., $\mathbf{G}_4=\mathbf{G}_5=\mathbf{G}_6$. Here, the rank of the coupling submatrix ($\mathbf{G}_4$, $\mathbf{G}_5$, $\mathbf{G}_6 $) is one. Therefore, there exist two dark states: one is given in Eq.~(\ref{4-2-1dark}), and the other is given in Eq.~(\ref{4-2-2dark}). Using the Gram-Schmidt orthogonalization, these two dark states can be orthogonalized as Eq.~(\ref{4-2-1dark}) and
	\begin{align}\label{4-2dark-orth}
	\vert \tilde{D}_{[4]}^{(2)}(2)\rangle
	&= \frac{1}{\sqrt{6}} \big( \vert L_{[4]}^{(2)}(4)\rangle
	+ \vert L_{[4]}^{(2)}(5)\rangle
	- 2 \vert L_{[4]}^{(2)}(6)\rangle \big) \nonumber \\
	&= \frac{\sqrt{3}}{6}\vert 0\rangle\big( -\vert e,e,g,g\rangle
	-\vert e,g,e,g\rangle
	+\vert g,e,g,e\rangle \big) \nonumber \\
	&\quad +\frac{2\sqrt{3}}{6}\vert 0\rangle\big( \vert e,g,g,e\rangle
	-\vert g,e,e,g\rangle \big) \nonumber \\
	&\quad +\frac{\sqrt{6}}{6}\vert 0\rangle\vert g,g,e,e\rangle.
	\end{align}
	These two states in Eq.~(\ref{4-2-1dark}) and Eq.~(\ref{4-2dark-orth}) form a complete set of orthogonal dark states for this degenerate subspace, and any unitary transformation of them can create a new set of orthogonal dark states.

	\subsubsection{Three-excitation subspace}
	
	In the three-excitation subspace, there are fifteen basis states which including eleven upper states $\{\vert u_{1}\rangle=\left\vert
	3,g,g,g,g\right\rangle $, $\vert u_{2}\rangle=\left\vert 2,e,g,g,g\right\rangle $, $\vert u_{3}\rangle=\left\vert
	2,g,e,g,g\right\rangle $, $\vert u_{4}\rangle=\left\vert 2,g,g,e,g\right\rangle $, $\vert u_{5}\rangle=\left\vert
	2,g,g,g,e\right\rangle $, $\vert u_{6}\rangle=\left\vert 1,e,e,g,g\right\rangle $, $\vert u_{7}\rangle=\left\vert
	1,e,g,e,g\right\rangle $, $\vert u_{8}\rangle=\left\vert 1,e,g,g,e\right\rangle $, $\vert u_{9}\rangle=\left\vert
	1,g,e,e,g\right\rangle $, $\vert u_{10}\rangle=\left\vert 1,g,e,g,e\right\rangle $, $\vert u_{11}\rangle=\left\vert
	1,g,g,e,e\right\rangle \}$ and four lower states $\{\vert l_1\rangle = \vert 0,e,e,e,g\rangle,\ 
	\vert l_2\rangle = \vert 0,e,e,g,e\rangle,\ 
	\vert l_3\rangle = \vert 0,e,g,e,e\rangle,\ 
	\vert l_4\rangle = \vert 0,g,e,e,e\rangle\}$. Similarly, we define the basis vectors in order and then the Hamiltonian can be written as
	\begin{equation}
		{H}_{[4]}^{\left( 3\right) }=\left(
		\begin{array}{c|c}
			\mathbf{{U}}_{\left[ 4\right] }^{\left( 3\right) } & \mathbf{{C}}%
			_{\left[ 4\right] }^{\left( 3\right) } \\ \hline
			\left( \mathbf{{C}}_{\left[ 4\right] }^{\left( 3\right) }\right)
			^{\dag } & \mathbf{{L}}_{\left[ 4\right] }^{\left( 3\right) }%
		\end{array}%
		\right) ,
	\end{equation}
	where the lower-state submatrix $\mathbf{L}_{[4]}^{[3]}$ and the coupling submatrix $\mathbf{{C}}_{\left[ 4\right] }^{\left( 3\right) }$
	can be expressed as%
	\begin{subequations}
		\begin{equation}
			\mathbf{L}_{[4]}^{(3)}=\left(
			\begin{array}{cccc}
				\Delta _{a} & V_{dd} & V_{dd} & V_{dd} \\
				V_{dd} & \Delta _{a} & V_{dd} & V_{dd} \\
				V_{dd} & V_{dd} & \Delta _{a} & V_{dd} \\
				V_{dd} & V_{dd} & V_{dd} & \Delta _{a}
			\end{array}
			\right) ,
		\end{equation}
		\begin{equation}
			\mathbf{C}_{[4]}^{(3)}=\left(
			\begin{array}{ccccccccccc}
				0 & 0 & 0 & 0 & 0 & g_{3} & g_{2} & 0 & g_{1} & 0 & 0 \\
				0 & 0 & 0 & 0 & 0 & g_{4} & 0 & g_{2} & 0 & g_{1} & 0 \\
				0 & 0 & 0 & 0 & 0 & 0 & g_{4} & g_{3} & 0 & 0 & g_{1} \\
				0 & 0 & 0 & 0 & 0 & 0 & 0 & 0 & g_{4} & g_{3} & g_{2}
			\end{array}
			\right) ^{T}.
		\end{equation}
	\end{subequations}
	
	The lower-state submatrix $\mathbf{L}_{[4]}^{[3]}$ has the same form as that in Eq.~(\ref{h4-1}), then it can be diagonalized with the unitary matrix in Eq.~(\ref{matrix-sl-4}). The diagonalized lower-state submatrix and the corresponding coupling submatrix are given by
\begin{subequations}\label{h4-3arrow}
	\begin{align}
		\mathbf{\tilde{L}}_{[4]}^{(3)} &= \operatorname{diag}\big(
		\Delta_{a}+3V_{dd},\Delta_{a}-V_{dd},\Delta_{a}-V_{dd},\Delta_{a}-V_{dd}\big), \\
		\mathbf{\tilde{C}}_{[4]}^{(3)} &= (\mathbf{G}_1, \mathbf{G}_2, \mathbf{G}_3, \mathbf{G}_4) \notag \\
		&= \begin{pmatrix}
			0 & 0 & 0 & 0 \\
			0 & 0 & 0 & 0 \\
			0 & 0 & 0 & 0 \\
			0 & 0 & 0 & 0 \\
			0 & 0 & 0 & 0 \\
			\frac{g_{3}+g_{4}}{2} & \frac{-g_{3}+g_{4}}{\sqrt{2}} & \frac{-g_{3}-g_{4}}{\sqrt{6}} & \frac{-g_{3}-g_{4}}{2\sqrt{3}} \\
			\frac{g_{2}+g_{4}}{2} & -\frac{g_{2}}{\sqrt{2}} & \frac{-g_{2}+2g_{4}}{\sqrt{6}} & \frac{-g_{2}-g_{4}}{2\sqrt{3}} \\
			\frac{g_{2}+g_{3}}{2} & \frac{g_{2}}{\sqrt{2}} & \frac{-g_{2}+2g_{3}}{\sqrt{6}} & \frac{-g_{2}-g_{3}}{2\sqrt{3}} \\
			\frac{g_{1}+g_{4}}{2} & \frac{-g_{1}}{\sqrt{2}} & \frac{-g_{1}}{\sqrt{6}} & \frac{-g_{1}+3g_{4}}{2\sqrt{3}} \\
			\frac{g_{1}+g_{3}}{2} & \frac{g_{1}}{\sqrt{2}} & \frac{-g_{1}}{\sqrt{6}} & \frac{-g_{1}+3g_{3}}{2\sqrt{3}} \\
			\frac{g_{1}+g_{2}}{2} & 0 & \frac{\sqrt{6}g_{1}}{3} & \frac{-g_{1}+3g_{2}}{2\sqrt{3}}
		\end{pmatrix},
	\end{align}
\end{subequations}
	and the dressed lower states of the Hamiltonian $\tilde{H}_{[4]}^{\left( 3\right)}$ as follows
	\begin{subequations}
		\begin{align}
			\vert L _{\left[4\right] }^{\left( 3\right) }\left(1\right)\rangle &= \frac{1}{2}\vert 0\rangle ( \vert e,e,e,g\rangle - \vert e,e,g,e\rangle - \vert e,g,e,e\rangle + \vert g,e,e,e\rangle),  \\
			\vert L _{\left[4\right] }^{\left( 3\right) }\left(2\right)\rangle &= \frac{1}{\sqrt{2}}\vert 0\rangle ( -\vert e,e,e,g\rangle + \vert g,e,e,e\rangle ),  \\
			\vert L _{\left[4\right] }^{\left( 3\right) }\left(3\right)\rangle &= \frac{1}{\sqrt{6}}\vert 0\rangle ( -\vert e,e,e,g\rangle + 2\vert e,g,e,e\rangle - \vert g,e,e,e\rangle),  \\
			\vert L _{\left[4\right] }^{\left( 3\right) }\left(4\right)\rangle &= \frac{\sqrt{3}}{6}\vert 0\rangle ( -\vert e,e,e,g\rangle + 2\vert e,e,g,e\rangle - \vert e,g,e,e\rangle \notag \\
			&\quad - \vert g,e,e,e\rangle ) .
		\end{align}
	\end{subequations}
	Here, the two states $\vert L _{\left[4\right] }^{\left( 3\right) }\left(1\right)\rangle$ and $\vert L _{\left[4\right] }^{\left( 3\right) }\left(4\right)\rangle$ are $W$-like states, which are entangled states involving four atoms. The state $\vert L _{\left[4\right] }^{\left( 3\right) }\left(2\right)\rangle$ is a Bell state involving the first and fourth atoms, while the second and third atoms are in the excited state $\vert e\rangle$. The state $\vert L _{\left[4\right] }^{\left( 3\right) }\left(3\right)\rangle$ is a $W$-like state involving the first, second, and fourth atoms, while the third atom is in the separate excited state $\vert e\rangle$.
	
	The dark states in this case can be obtained by analyzing the Hamiltonian $\tilde{H}_{[4]}^{\left( 3\right) }$ based on Eq.~(\ref{h4-3arrow}) with the arrowhead-matrix method.
	 
	(1) Consider the case of zero coupling column vector: For non-zero $g_1$, $g_2$, $g_3$ and $g_4$, there is no zero coupling vector, then there is no dark state related to the zero coupling vector. 
	
	(2) Consider the case of degenerate lower-state subspace: 
	There is a three-dimensional degenerate subspace $\{\vert L _{\left[4\right] }^{\left( 3\right) }\left(2\right)\rangle ,\vert L _{\left[4\right] }^{\left( 3\right) }\left(3\right)\rangle,\vert L _{\left[4\right] }^{\left( 3\right) }\left(4\right)\rangle \}$, and the corresponding coupling submatrix is full rank with non-zero $g_j$ (for $j=1$-$4$). As a result, there is no dark state. 
	
	\begin{figure}[t!]
		\centering\includegraphics[width=0.48\textwidth]{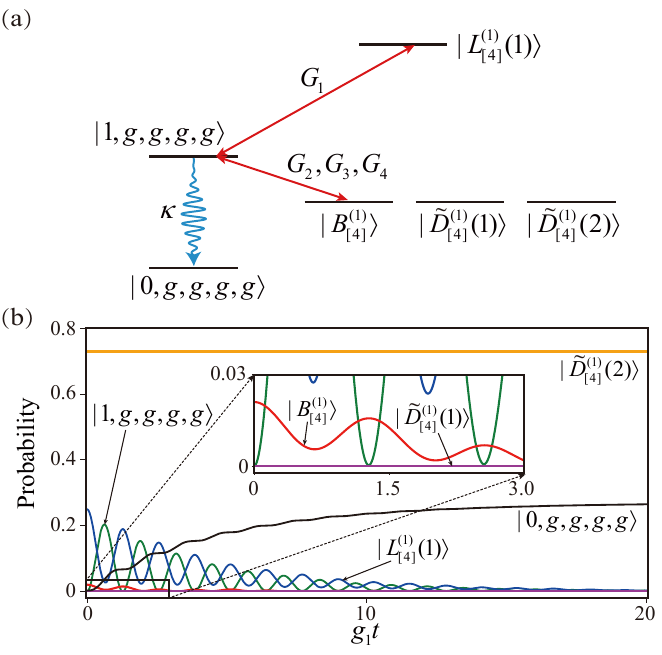}  
		\caption{(a) Energy-level diagram of the coupled cavity-four-atom system confined in the zero- and single-excitation subspaces. (b) The occupation probabilities of these states $\vert 1,g,g,g,g \rangle $ (green), $\vert L _{\left[4\right] }^{\left( 1\right) }\left(1\right)\rangle $ (blue), $\vert B _{\left[4\right] }^{\left( 1\right) }\rangle$ (red), $\vert \tilde{D} _{\left[4\right] }^{\left( 1\right) }\left(1\right)\rangle $ (purple), $\vert \tilde{D} _{\left[4\right] }^{\left( 1\right) }\left(2\right)\rangle $ (yellow), and $\vert 0,g,g,g,g \rangle $ (black) as functions of time in the zero- and single-excitation subspaces case, when the initial state of the system is $\vert 0,g,g,g,e \rangle $. Other parameters used are $g_{2}/g_{1}=0.8$, $g_{3}/g_{1}=1.5$, $g_{4}/g_{1}=1.2$, $V_{dd}/g_{1}=0.5$, $\Delta _{a}/g_{1}= 0$ and $\kappa/g_{1}=0.5$.}
		\label{four-open}
	\end{figure}
	
	\subsection{Characterization of the dark states}
	
	 In the open-system case, the dynamical evolution of the system is governed by the quantum master equation [$N=4$ for Eq.~(\ref{master})]. As we studied in the above section, the dark-state effect appears in the single- and double-excitation subspaces for the four-atom case. Here we present the characterization of the dark states in both the single-excitation and double-excitation subspaces. In particular, for the double-excitation case, there are several groups of dark states depending on the parameter conditions. For simplicity, we only consider the case of $\mathbf{G}_4=\mathbf{G}_5$ as an example. 
	 
	 Below, we first study the dark-state characterization in the single-excitation subspace. To this end, we analyze the energy-level transitions of the system. For simplicity, here we only show the energy levels of the system within the ground state and single-excitation subspaces. This is reasonable because there is no driving and the environment is a vacuum bath, and hence there is no exciting process in the system.
	 
	 In Fig.~\ref{four-open}(a), we plot the energy-level diagram of the four-atom case in the zero- and single-excitation subspaces. In this case, the dressed lower state $\vert L _{\left[4\right] }^{\left( 1\right) }\left(1\right)\rangle$ and the bright state $\vert B _{\left[4\right] }^{\left( 1\right) }\rangle =( G_{2}\vert L _{\left[4\right] }^{\left( 1\right) }\left(2\right)\rangle + G_{3}\vert L _{\left[4\right] }^{\left( 1\right) }\left(3\right)\rangle+ G_{4}\vert L _{\left[4\right] }^{\left( 1\right) }\left(4\right)\rangle)/(G_{2}^{2}+G_{3}^{2}+G_{4}^{2})^{1/2}$ are coupled to the upper state $\vert 1,g,g,g,g \rangle$. In addition, cavity-field dissipation provides a decay pathway from the upper state $\vert 1,g,g,g,g \rangle$ to the ground state $\vert 0,g,g,g,g \rangle$. In contrast, the dark states $\vert \tilde{D} _{\left[4\right] }^{\left( 1\right) }(1)\rangle$ and $\vert \tilde{D} _{\left[4\right] }^{\left( 1\right) }(2)\rangle$ are decoupled from the upper state, then they are immune to the dissipative processes of cavity. Similarly, we find that when the initial state is $\vert 0,g,g,g,e \rangle $, the population of the dark state $\vert \tilde{D} _{\left[4\right] }^{\left( 1\right) }(1)\rangle$ is zero, because the project of the state $\vert \tilde{D} _{\left[4\right] }^{\left( 1\right) }(1)\rangle$ onto state $\vert 0,g,g,g,e \rangle $ is zero. The initial state $\vert 0,g,g,g,e \rangle $ can be expressed as a superposition of the these states $\vert L _{\left[4\right] }^{\left( 1\right) }\left(1\right)\rangle$, $\vert B _{\left[4\right] }^{\left( 1\right) }\rangle$, and $\vert \tilde{D} _{\left[4\right] }^{\left( 1\right) }(2)\rangle$, and then the system evolves toward a steady state that is a superposition of the dark state $\vert \tilde{D} _{\left[4\right] }^{\left( 1\right) }(2)\rangle$ and the ground state $\vert 0,g,g,g,g \rangle $ because of the dissipation of the cavity. Therefore, a finite steady-state population serves as the manifestation of the dark state $\vert \tilde{D} _{\left[4\right] }^{\left( 1\right) }(2)\rangle$, offering a practical signature for identifying dark-state effect. Notably, by measuring the excited-state probabilities of the three atoms, the dark-state population can be distinguished from the ground state $|0,g,g,g,g\rangle$. Hence, the proposed method for characterizing dark states is feasible for experimental implementation.
	 
	 In Fig.~\ref{four-open}(b), we plot the populations of these states $\vert 1,g,g,g,g \rangle $, $\vert L _{\left[4\right] }^{\left( 1\right) }\left(1\right)\rangle $, $\vert B _{\left[4\right] }^{\left( 1\right) }\rangle$, $\vert \tilde{D} _{\left[4\right] }^{\left( 1\right) }\left(1\right)\rangle $, $\vert \tilde{D} _{\left[4\right] }^{\left( 1\right) }\left(2\right)\rangle $, and $\vert 0,g,g,g,g \rangle $ as functions of the scaled time $g_1t$ when the initial state is $\vert 0,g,g,g,e \rangle$. This initial state can be prepared by only driving the fourth atom with the Hamiltonian in Eq.~(\ref{qudong-a}). At the initial time, populations are distributed among these three states  $\vert L _{\left[4\right] }^{\left( 1\right) }\left(1\right)\rangle$, $\vert B _{\left[4\right] }^{\left( 1\right) }\rangle$, and $\vert \tilde{D} _{\left[4\right] }^{\left( 1\right) }(2)\rangle$, while the population of the dark state $\vert \tilde{D} _{\left[4\right] }^{\left( 1\right) }(1)\rangle$ is zero. As time evolves, the populations of these states $\vert 1,g,g,g,g \rangle$, $\vert L _{\left[4\right] }^{\left( 1\right) }\left(1\right)\rangle$, and $\vert B _{\left[4\right] }^{\left( 1\right) }\rangle$ exhibit oscillations, then decay to the ground state $\vert 0,g,g,g,g \rangle$, while the population of the dark state $\vert \tilde{D} _{\left[4\right] }^{\left( 1\right) }(2)\rangle$ stays constant for all times. Therefore, the system eventually relaxes to a steady state with population only in the dark state $\vert \tilde{D} _{\left[4\right] }^{\left( 1\right) }(2)\rangle$ and the ground state $\vert 0,g,g,g,g \rangle$. Note that these two states exhibit distinct cases in atomic populations, therefore, we can distinguish these two states by measuring the atomic populations.
	 
	 \begin{figure}[t!]
	 	\centering\includegraphics[width=0.48\textwidth]{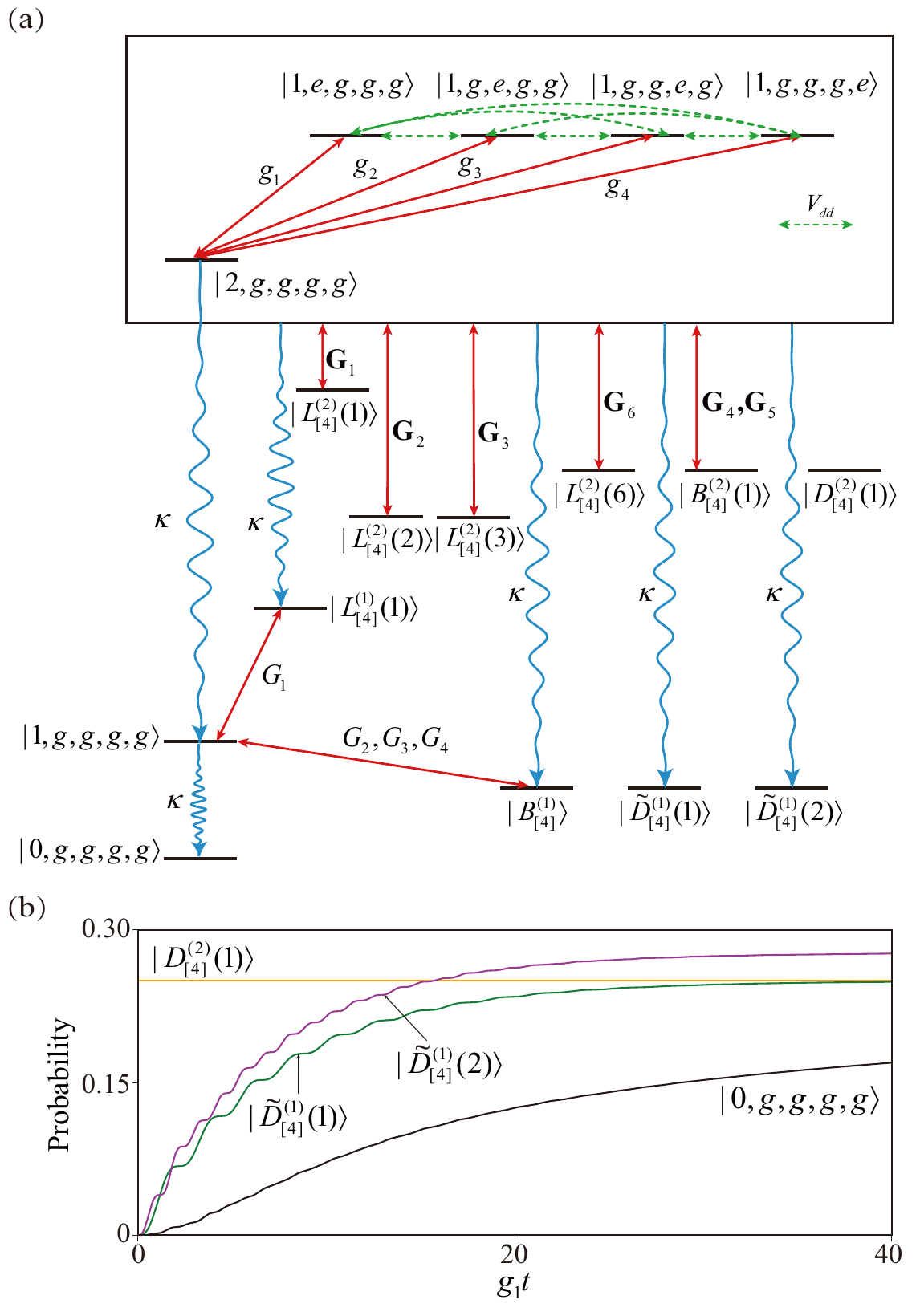}  
	 	\caption{(a) Energy-level diagram of the coupled cavity-four-atom system confined in the zero-, single-, and double-excitation subspaces. (b) The occupation probabilities of these states $\vert D _{\left[4\right] }^{\left( 2\right) }\left(1\right)\rangle $ (yellow), $\vert \tilde{D} _{\left[4\right] }^{\left( 1\right) }\left(1\right)\rangle $ (green), $\vert \tilde{D} _{\left[4\right] }^{\left( 1\right) }\left(2\right)\rangle $ (purple), and $\vert 0,g,g,g,g \rangle $ (black) as functions of time in the zero-, single, and double-excitation subspaces case, when the initial state of the system is $\vert 0,e,e,g,g \rangle $. Other parameters used are $g_{2}/g_{1}=2$, $g_{3}/g_{1}=2$, $g_{4}/g_{1}=-1$, $V_{dd}/g_{1}=0.5$, $\Delta _{a}/g_{1}= 0$, and $\kappa/g_{1}=0.3$.}
	 	\label{four-open-two}
	 \end{figure}
	 
	 We now turn to the dark-state characterization in the double-excitation subspaces. We first analyze the energy-level transitions in this system. Similarly, the transitions only involve these states in the double-, single-, and zero-excitation subspaces, because these is no driving and the environment is a vacuum bath. In the double- and single-excitation subspaces, there are eleven and five basis states, respectively. In addition, there is a single ground state. In Fig.~\ref{four-open-two}(a), we show the transitions among these subspaces and states. We can see that the cavity-field dissipation will induce decay to connect different excitation subspaces (only from high-excitation subspace to lower-excitation subspace), corresponding to these transitions labeled by $\kappa$. In particular, the cavity-field decay will induce the transitions from these upper states in the double-excitation subspace to these lower states in the single-excitation subspace. These transitions lead to the stead-state populations for the dark states in the single-excitation subspace, when the system is initially in the double-excitation state. Meanwhile, the cavity decay connects these states $\left\vert 2,g,g,g,g\right\rangle \rightarrow \left\vert1,g,g,g,g\right\rangle \rightarrow \left\vert 0,g,g,g,g\right\rangle $. In addition, the dark states will be decoupled from the upper states in the same subspace. The coherent coupling (the atom-field coupling and atom-atom coupling) will induce transitions within the same subspaces  (these transitions labeled by $G_{1\text{-}4}$ and $\mathbf{G}_{1\text{-}6}$).
	 
	 In Fig.~\ref{four-open-two}(b), we plot the occupation probabilities of these states $\vert D _{\left[4\right] }^{\left( 2\right) }\left(1\right)\rangle $, $\vert \tilde{D} _{\left[4\right] }^{\left( 1\right) }\left(1\right)\rangle $, $\vert \tilde{D} _{\left[4\right] }^{\left( 1\right) }\left(2\right)\rangle $, and $\vert 0,g,g,g,g \rangle $ as functions of the scaled time $g_1t$ in the zero-, single-, and double-excitation subspaces, when the initial state of the system is $\vert 0,e,e,g,g \rangle $. This initial state can be prepared by driving the first and second atoms with the Hamiltonian in Eq.~(\ref{qudong-a}). At the initial time, the system will be in a superposition of these states: $\vert L _{\left[4\right] }^{\left( 2\right) }\left(1\right)\rangle $, $\vert L _{\left[4\right] }^{\left( 2\right) }\left(2\right)\rangle $, $\vert L _{\left[4\right] }^{\left( 2\right) }\left(3\right)\rangle $, and $\vert L _{\left[4\right] }^{\left( 2\right) }\left(4\right)\rangle $. Under the condition $\mathbf{G}_4=\mathbf{G}_5$, the proper superpositions of $\vert L _{\left[4\right] }^{\left( 2\right) }\left(4\right)\rangle $ and $\vert L _{\left[4\right] }^{\left( 2\right) }\left(5\right)\rangle $ will form the dark state $\vert D _{\left[4\right] }^{\left( 2\right) }\left(1\right)\rangle $ and a bright state $\vert B _{\left[4\right] }^{\left( 2\right) }\left(1\right)\rangle $. Corresponding to the initial state $\vert 0,e,e,g,g \rangle $, the system will initially be in a superposition of the above mention states. As a result, the system will be partially populated in the dark state $\vert D _{\left[4\right] }^{\left( 2\right) }\left(1\right)\rangle $, while the population of the dark state $\vert \tilde{D} _{\left[4\right] }^{\left( 1\right) }\left(1\right)\rangle $ and $\vert \tilde{D} _{\left[4\right] }^{\left( 1\right) }\left(2\right)\rangle $ is zero, since the single-excitation subspace is not initially populated. As time evolves, the populations of these three states $\vert \tilde{D} _{\left[4\right] }^{\left( 1\right) }\left(1\right)\rangle $, $\vert \tilde{D} _{\left[4\right] }^{\left( 1\right) }\left(2\right)\rangle $, and $\vert 0,g,g,g,g \rangle $ present an upward trend (the physical reason has been explained in the above paragraph), while the population of the dark state $\vert D _{\left[4\right] }^{\left( 2\right) }\left(1\right)\rangle $ remains constant in time. This dynamics can be understood based on the transitions in the system. Therefore, the system eventually relaxes to a steady state with population in the dark states $\vert D _{\left[4\right] }^{\left( 2\right) }\left(1\right)\rangle $, $\vert \tilde{D} _{\left[4\right] }^{\left( 1\right) }\left(1\right)\rangle $, $\vert \tilde{D} _{\left[4\right] }^{\left( 1\right) }\left(2\right)\rangle $, and the ground state $\vert 0,g,g,g,g \rangle $. Then we can identify the dark states $\vert D _{\left[4\right] }^{\left( 2\right) }\left(1\right)\rangle $ by measuring the atomic populations, and there exist two excited-state populations in these four atoms. Note that all these four states shown in Fig.~\ref{four-open}(c) have no photon component, then these four states in different excitation-number subspaces can be distinguished based on the excitations. Therefore, the dark state in the two-excitation subspace can be characterized from the population dynamics.
	
	\section{Dark states in the $\boldsymbol{N}$-atom case}\label{sec6}
	
	In this section, we apply the arrowhead-matrix method to find the dark states when $N$ Rydberg atoms are coupled to the cavity field, which is described by the Hamiltonian $%
	\hat{H}_{[N]}$ in Eq.~(\ref{Hn}).
	Theoretically, the dark-state effect of the system in a general excitation-number subspace can be analyzed with the arrowhead-matrix method. However, for large values of $N$ and $n$, it is not a trivial task to obtain the analytical results of the dark states. Below, we present some analytical results concerning the dark states in the single-excitation subspace ($n=1$) for a general coupled cavity-$N$-atom system.
	
	In the single-excitation subspace of the $N$-atom system, the basis states are given by $\left\{
	\left\vert 1,g,g,\dots,g\right\rangle ,\left\vert 0,e,g,\dots,g\right\rangle
	,\dots,\left\vert 0,g,g,\dots,e\right\rangle \right\}$, and there is one upper
	state associated with the cavity-field number state $\left\vert 1\right\rangle $
	and $N$ lower states associated with the vacuum state $\left\vert
	0\right\rangle .$ We define the basis vectors as $\vert u_{1}\rangle=\left\vert
	1,g,g,\dots,g\right\rangle =\left( 1,0,0,\dots,0\right) ^{T}$, $\vert l_{1}\rangle=\left\vert
	0,e,g,\dots,g\right\rangle =\left( 0,1,0,\dots,0\right) ^{T}$, $\ldots$, and $\vert l_{N}\rangle=\left\vert 0,g,g,\dots,e\right\rangle =\left( 0,0,0,\dots,1\right) ^{T},$ then
	the Hamiltonian $\hat{H}_{[N]}$ in the single-excitation subspace can be expressed
	as%
	\begin{equation}
		H_{[N]}^{\left( 1\right) }=\left(
		\begin{array}{c|cccc}
			-\frac{N\Delta _{a}}{2} & g_{1} & g_{2} & \dots & g_{N} \\ \hline
			g_{1} & -\frac{\left( N-2\right) \Delta _{a}}{2} & V_{dd} & \cdots & V_{dd}\\
			g_{2} & V_{dd} & -\frac{\left( N-2\right) \Delta _{a}}{2} & \cdots & V_{dd}\\
			\vdots & \vdots & \vdots & \ddots & \vdots \\
			g_{N} & V_{dd} & V_{dd} & \cdots & -\frac{\left( N-2\right) \Delta _{a}}{2}%
		\end{array}%
		\right) .
	\end{equation}
	Based on the previous analyses on the two-, three-, and four-atom cases, we find that the lower-state submatrix $\mathbf{L}_{[N]}^{(1)}$ can always be diagonalized with the unitary matrix 
	\begin{equation}
		\mathbf{S}_{l}=\left(
		\begin{array}{ccccc}
			\frac{1}{\sqrt{N}} & \frac{1}{\sqrt{N}} & \frac{1}{\sqrt{N}} & \dots & \frac{1}{\sqrt{N}} \\ 
			-\frac{1}{\sqrt{2}} & \frac{1}{\sqrt{2}} & 0 & \cdots & 0\\
			-\frac{1}{\sqrt{6}} & -\frac{1}{\sqrt{6}} & \frac{2}{\sqrt{6}} & \cdots & 0\\
			\vdots & \vdots & \vdots & \ddots & \vdots \\
			-\frac{1}{\sqrt{N\left( N-1\right) }} & -\frac{1}{\sqrt{N\left( N-1\right) }} & -\frac{1}{\sqrt{N\left( N-1\right) }} & \cdots & \frac{N-1}{\sqrt{N\left(
					N-1\right) }}%
		\end{array}%
		\right) .
	\end{equation}
	Then the transformed Hamiltonian can be expressed as an arrowhead matrix
	\begin{equation}\label{hn-1-arrow}
		\tilde{H}_{[N]}^{\left( 1\right) }=\left(
		\begin{array}{c|cccc}
			-\frac{N\Delta _{a}}{2} & G_{1} & G_{2} & \dots & G_{N} \\ \hline
			G_{1} & \lambda_{0} & 0 & \cdots & 0\\
			G_{2} & 0 & \lambda_{1} & \cdots & 0\\
			\vdots & \vdots & \vdots & \ddots & \vdots \\
			G_{N} & 0 & 0 & \cdots & \lambda_{1}%
		\end{array}%
		\right) ,
	\end{equation}
	where the parameters $\lambda_{0}=-{\left( N-2\right) \Delta _{a}}/{2}+(N-1)V_{dd}$ and $\lambda_{1}=-{\left( N-2\right) \Delta _{a}}/{2}-V_{dd}$ are introduced. The corresponding coupling matrix can be obtained as $\mathbf{\tilde{C}}_{[N]}^{\left( 1\right) }=\mathbf{C}_{[N]}^{\left( 1\right) }\mathbf{S_{l}}^{\dag
	}=\left( G_{1},G_{2},\dots,G_{N}\right) $, where we introduce the coupling strengths in the arrowhead matrix,
	\begin{subequations}\label{Gj}
		\begin{align}
			G_{1} &= \frac{1}{\sqrt{N}}\sum_{j^{\prime }=1}^{N}g_{j^{\prime }}, \\
			G_{s} &=\sum_{l=1}^{s-1}\frac{-1}{\sqrt{l\left( l-1\right) }}g_{l}
			+\frac{s-1}{\sqrt{s\left( s-1\right) }}g_{s},
		\end{align}
	\end{subequations}
	for $s=2,3,\dots,N$ and $l=1,2,\dots,s-1$. The corresponding dressed lower states are given by
	\begin{subequations}\label{Lj}
		\begin{align}
			\vert L _{\left[N\right] }^{\left( 1\right) }\left(1\right)\rangle
			&= \frac{1}{\sqrt{N}}\sum_{j'=1}^{N}\vert l_{j'}\rangle, \\
			\vert L _{\left[N\right] }^{\left( 1\right) }\left(s\right)\rangle
			&= \sum_{l=1}^{s-1}\frac{-1}{\sqrt{l(l-1)}}\left\vert l_{l}\right\rangle
			+ \frac{s-1}{\sqrt{s(s-1)}}\left\vert l_{s}\right\rangle.
		\end{align}
	\end{subequations}
	The dark states in this case can be obtained by analyzing Eq.~(\ref{hn-1-arrow}) with the arrowhead-matrix method.
	
	(1) Consider the case of zero coupling column vector:
	When $G_{j}=0$ ($j=1,2,\dots,N$), the corresponding dressed lower state $\vert L_{j}\rangle$ is decoupled from the upper state $\vert u_{1}\rangle$ and becomes a dark state.
	
	(2) Consider the case of degenerate lower-state subspace:
	It can be seen from Eq.~(\ref{hn-1-arrow}) that there are $N-1$ degenerate values $\lambda_{1}$, then there is a ($N-1$)-dimensional degenerate lower-state subspace $\{\vert L _{\left[N\right] }^{\left( 1\right) }\left(2\right)\rangle ,\vert L _{\left[N\right] }^{\left( 1\right) }\left(3\right)\rangle,\dots,\vert L _{\left[N\right] }^{\left( 1\right) }\left(N\right)\rangle \}$. As a result, there are $N-2$ dark states
	\begin{subequations}
		\begin{align}
			\vert D_{[N]}^{( 1) }\left(1\right) \rangle &= \frac{G_{3}\vert L _{[N] }^{( 1) }\left(2\right)\rangle -G_{2}\vert L _{\left[N\right] }^{\left( 1\right) }\left(3\right)\rangle}{\sqrt{G_{3}^{2}+G_{2}^{2}}}, \\
			\vert D_{[N]}^{\left( 1\right) }\left( 2\right) \rangle &= \frac{G_{4}\vert L _{\left[N\right] }^{\left( 1\right) }\left(2\right)\rangle -G_{2}\vert L _{\left[N\right] }^{\left( 1\right) }\left(4\right)\rangle}{\sqrt{G_{4}^{2}+G_{2}^{2}}}, \\
			& \quad \dots \nonumber \\
			\vert D_{[N]}^{\left( 1\right) }\left( l\right) \rangle &= \frac{G_{l+2}\vert L _{\left[N\right] }^{\left( 1\right) }\left(2\right)\rangle -G_{2}\vert L _{\left[N\right] }^{\left( 1\right) }\left(l+2\right)\rangle}{\sqrt{G_{l+2}^{2}+G_{2}^{2}}}, \\
			& \quad \dots \nonumber \\
			\vert D_{[N]}^{\left( 1\right) }\left( N-2\right) \rangle &= \frac{G_{N}\vert L _{\left[N\right] }^{\left( 1\right) }\left(2\right)\rangle -G_{2}\vert L _{\left[N\right] }^{\left( 1\right) }\left(N\right)\rangle}{\sqrt{G_{N}^{2}+G_{2}^{2}}},
		\end{align}
	\end{subequations}
	where $l=1,2$, $\dots$, $N-2$. Note that the form of the dark states are not unique. With the Gram-Schmidt orthogonalization, we can obtain the orthogonalized dark states as
	\begin{subequations}
		\begin{align}
			\vert \tilde{D}_{[N]}^{(1)}\left(1\right) \rangle &= 
			\frac{G_{3}\vert L _{[N] }^{(1) }\left(2\right)\rangle - G_{2}\vert L _{[N] }^{( 1) }\left(3\right)\rangle}{\sqrt{G_{3}^{2}+G_{2}^{2}}}, \\
			\vert \tilde{D}_{[N]}^{(1)}\left(2\right) \rangle &= 
			\frac{1}{\sqrt{(G_{3}^{2}+G_{2}^{2})(G_{4}^{2}+G_{3}^{2}+G_{2}^{2})}}\notag \\
			& \quad \times \!\big[G_{2}G_{4}\vert L _{[N] }^{( 1) }\left(2\right)\rangle + G_{3}G_{4}\vert L _{[N] }^{( 1) }\left(3\right)\rangle \notag \\
			& \quad- (G_{2}^{2}+G_{3}^{2})\vert L _{[N] }^{( 1) }\left(4\right)\rangle\bigr], \\
			& \quad \dots \nonumber \\
			\vert \tilde{D}_{[N]}^{(1)}\left(l\right) \rangle &= 
			\frac{1}{\sqrt{(G_{2}^{2}+\dots+G_{l+1}^{2})(G_{2}^{2}+\dots+G_{l+2}^{2})}} \notag \\
			& \quad \times \! \big[ G_{2}G_{l+2}\vert L _{\left[N\right] }^{\left( 1\right) }\left(2\right)\rangle +\sum_{j=3}^{l+1}G_{j}G_{l+2}\vert L _{[N] }^{( 1) }\left(j\right)\rangle \notag \\
			& \quad - (G_{2}^{2}+G_{3}^{2}+\dots+G_{l+1}^{2})\vert L _{\left[N\right] }^{\left( 1\right) }\left(l+2\right)\rangle \bigr] , \\
			& \quad \dots \nonumber \\
			\vert \tilde{D}_{[N]}^{(1)}\left(N-2\right) \rangle &=
			\frac{1}{\sqrt{(G_{2}^{2}+\dots+G_{N-1}^{2})(G_{2}^{2}+\dots+G_{N}^{2})}} \notag \\
			& \quad \times \! \bigl[ G_{2}G_{N}\vert L _{[N] }^{( 1) }\left(2\right)\rangle +\sum_{j=3}^{N-1}G_{j}G_{N}\vert L _{[N] }^{( 1) }\left(j\right)\rangle \notag \\
			& \quad - (G_{2}^{2}+\dots+G_{N-1}^{2})\vert L _{[N] }^{( 1) }\left(N\right)\rangle \bigr] .
		\end{align}
	\end{subequations}
	
	In the $n$-excitation subspace $\left( n< N\right) $, the numbers of the upper
	and lower states can be represented by combination number: $1+C_{N}^{1}+C_{N}^{2}+\dots+C_{N}^{n-1}$ and $C_{N}^{n}$, where $C_{N}^{n}=N!/[n!\left( N-n\right) !]$ and $!$ stands for factorial. Then by diagonalizing
	the lower-state submatrix in the transformed arrowhead matrix and according to the dark-state theorems, we can
	obtain the corresponding dark states in the $N$-atom case.
	When $n \geq N$, the system has no lower state and dark state.

	\section{Dark states in the general-interaction case}\label{sec7}
	
	In the above discussions, we have considered constant dipole-dipole couplings between these atoms for simplicity. In a realistic case, the coupling strength of the dipole-dipole interaction between two Rydberg atoms is given by $V_{jk}=C_{3}/R_{jk}^{3}$, where $C_{3}$ is the dipole-dipole
	interaction strength parameter, and $R_{jk}$ is the distance between the $j$th and $k$th atoms. We denote the position of these atoms as $\vec{r}_{j}=( x_{j},y_{j},z_{j}) $, then the dipole-dipole coupling strength can be obtained as $V_{jk}=C_{3}/[( x_{j}-x_{k}) ^{2}+( y_{j}-y_{k}) ^{2}+( z_{j}-z_{k}) ^{2}]^{{3}/{2}}$. As a result, the atomic Hamiltonian of the system can be expressed as
	\begin{equation}
		\hat{H}_{\text{atom}}=\frac{\omega _{a}}{2}\sum_{j=1}^{N}\hat{\sigma} _{j}^{z}+\sum_{j<k}\frac{%
			C_{3}}{R_{jk} ^{3}}( \hat{\sigma} _{j}^{+}\hat{\sigma} _{k}^{-}+%
		\text{H.c.}) .
	\end{equation}%
	Meanwhile, the coupling strength $g_j$ between the cavity field and the $j$th atom depend on the field distribution of the cavity in a realistic case. Typically, we consider a stand wave with a transverse Gaussian distribution, then the coupling strength between the $j$th atom and the field mode can be expressed as~\cite{walls}
	\begin{equation}
		g_j( \vec{r}_{j}) =g_{0}\text{cos}( kz_j) \exp \left( -\frac{%
			x_j^{2}+y_j^{2}}{w_{0}^{2}}\right) \left( \frac{w_{0}}{w _{z_j}}%
		\right),
	\end{equation}
	where $w_{0}$ is the beam waist radius, $w_{z_j}=w_{0}\sqrt{1+ z_j/z_{R}}$ is the beam width at position $z_j$ with the Rayleigh length $z_{R}=\pi w_{0}^{2}/\lambda$, and $\lambda$ is wavelength of the cavity field. Under the detailed $V_{jk}$ and $g_j(\vec{r}_{j})$, we can obtain a realistic physical model for the coupled cavity-atom system.
	
	Below, we study the dark-state effect in the realistic physical model. For simplicity, we first consider the two-atom case. Similar to the identical coupling case, we can analyze the dark-state effect in different excitation-number subspaces by rewriting the corresponding matrix under the replacement of $g\rightarrow g_{j}( \vec{r}_{j}) $ and $V_{dd} \rightarrow V_{jk} $. In the single-excitation subspace, the Hamiltonian $\hat{H}_{[2]}$ in realistic case can be rewritten from Eq.~(\ref{H21}) as %
	\begin{equation}
		H_{[2]}^{(1)}(\vec{r}_{1},\vec{r}_{2})=\left(
		\begin{array}{c|cc}
			-\Delta _{a} & g_{1}(\vec{r}_{1}) & g_{2}(\vec{r}_{2}) \\ \hline
			g_{1}(\vec{r}_{1}) & 0 & V_{12} \\
			g_{2}(\vec{r}_{2}) & V_{12} & 0%
		\end{array}%
		\right) .  \label{H-real-21}
	\end{equation} 
	By diagonalizing the lower-state submatrix with the unitary matrix in Eq.~(\ref{sl2-1}), then the Hamiltonian can be transformed into an arrowhead matrix
	\begin{equation}\label{h-real-2-1arrow}
		\tilde{H}_{[2]}^{( 1)}(\vec{r}_{1},\vec{r}_{2})=\left(
		\begin{array}{c|cc}
			-\Delta _{a} & G_{1}(\vec{r}_{1},\vec{r}_{2}) &G_{2}(\vec{r}_{1},\vec{r}_{2}) 
			\\ \hline
			G_{1}(\vec{r}_{1},\vec{r}_{2})& V_{12} & 0 \\
			G_{2}(\vec{r}_{1},\vec{r}_{2})& 0 & -V_{12}%
		\end{array}%
		\right) ,
	\end{equation}
	where the coupling strengths are introduced as $G_{1}(\vec{r}_{1},\vec{r}_{2})=[{g_{1}(\vec{r}_{1})+g_{2}(\vec{r}_{2})}]/{\sqrt{2}}$ and $G_{2}(\vec{r}_{1},\vec{r}_{2})=[{-g_{1}(\vec{r}_{1})+g_{2}(\vec{r}_{2})}]/{\sqrt{2}}$. Here, these three new basis states for the Hamiltonian $\tilde{H}_{[2]}^{\left( 1\right)}(\vec{r}_{1},\vec{r}_{2})$ are given in Eq.~(\ref{two-state}).
	
	The dark states in this case can be obtained by analyzing Eq.~(\ref{h-real-2-1arrow}) with the arrowhead-matrix method.
	
	(1) Consider the case of zero coupling column vector:
	(i) When $g_{1}(\vec{r}_{1})=-g_{2}(\vec{r}_{2})$, the corresponding coupling strength $G_{1}(\vec{r}_{1},\vec{r}_{2})$ between the lower state $\vert L _{\left[2\right] }^{\left( 1\right) }\left(1\right)\rangle$ and the upper state $\vert u_{1}\rangle$ is zero and then $\vert L _{\left[2\right] }^{\left( 1\right) }\left(1\right)\rangle$ becomes a dark state.
	(ii) When $g_{1}(\vec{r}_{1})=g_{2}(\vec{r}_{2})$, we have $G_{2}(\vec{r}_{1},\vec{r}_{2})=0$, then the state $\vert L _{\left[2\right] }^{\left( 1\right) }\left(2\right)\rangle$ is decoupled from the upper state $\vert u_{1}\rangle$ and becomes a dark state.
	
	(2) Consider the case of degenerate lower-state subspace: In a realistic case, we consider the dipole-dipole interaction strength $V_{12}\neq 0$, then there is no degeneracy in the lower states.
	
	It can be found that the results of the dark states are consistent with these we obtained in previous sections. Therefore, the dark states can be characterized in the same way.
	
	We now turn to the three-atom case. For simplicity, here we only consider the dark state in the single-excitation subspace. Similarly, the Hamiltonian $\hat{H}_{[3]}$ in Eq.~(\ref{h3-1}) can be rewritten as%
	\begin{equation}\label{h-real-3-1}
		H_{[3]}^{\left( 1\right) }(\vec{r}_{1},\vec{r}_{2},\vec{r}_{3})=\left(
		\begin{array}{c|ccc}
			-\frac{3}{2}\Delta _{a} & g_{1}(\vec{r}_{1}) & g_{2}(\vec{r}_{2}) & g_{3}(\vec{r}_{3}) \\ \hline
			g_{1}(\vec{r}_{1}) & -\frac{1}{2}\Delta _{a} & V_{12} & V_{13} \\
			g_{2}(\vec{r}_{2}) & V_{12} & -\frac{1}{2}\Delta _{a} & V_{23} \\
			g_{3}(\vec{r}_{3}) & V_{13} & V_{23} & -\frac{1}{2}\Delta _{a}%
		\end{array}%
		\right) .
	\end{equation}
	To analyze the dark states, we need to transform the Hamiltonian matrix in Eq.~(\ref{h-real-3-1}) into an arrowhead matrix. Next we analyze the case where the lower-state submatrix exhibits degeneracy. The lower-state submatriax can be expressed as $M=-{\Delta _{a}}I_{3}/{2}+V$, where the matrices $I_{3}$ and $V$ are introduced as
	\begin{subequations}
		\begin{align}
			I_3 &=
			\begin{pmatrix}
				1 & 0 & 0 \\
				0 & 1 & 0 \\
				0 & 0 & 1
			\end{pmatrix}, \\
			V &=
			\begin{pmatrix}
				0 & V_{12} & V_{13} \\
				V_{12} & 0 & V_{23} \\
				V_{13} & V_{23} & 0
			\end{pmatrix}.
		\end{align}
	\end{subequations}
	We assume that $\lambda$ is the eigenvalue of the matrix $V$, then the characteristic equation reads
	\begin{equation}\label{eq-tezhengzhi}
		\lambda ^{3}+P\lambda +Q=0,
	\end{equation}
	where these parameters are introduced as $P=-(V_{12}^{2}+V_{13}^{2}+V_{23}^{2})$ and $Q=-2V_{12}V_{13}V_{23}$. According to Cardano's formula~\cite{kadan}, the discriminant corresponding to Eq.~(\ref{eq-tezhengzhi}) is given by
	\begin{equation}
		\Delta =\left( \frac{P}{3}\right) ^{3}+\left( \frac{Q}{2}\right) ^{2}.
	\end{equation}
	When $\Delta=0$, Eq.~(\ref{eq-tezhengzhi}) possesses a repeated root. In the case of a double root, the condition $\left\vert V_{12}\right\vert = \left\vert V_{13}\right\vert=\left\vert V_{23}\right\vert $ must be satisfied. In the case of a triple root, the system need satisfy the condition $V_{12}=V_{13}=V_{23}=0$. For a realistic system, we consider $V_{jk}\neq 0$. Therefore, the dark states exist when the atomic interaction strength satisfy the condition $\left\vert V_{12}\right\vert = \left\vert V_{13}\right\vert=\left\vert V_{23}\right\vert $.
	
	When $\left\vert V_{12}\right\vert = \left\vert V_{13}\right\vert=\left\vert V_{23}\right\vert=V_{dd} $, the Hamiltonian in Eq.~(\ref{h-real-3-1}) becomes
	\begin{equation}\label{h-real-3-1-vdd}
		H_{[3]}^{\left( 1\right) }(\vec{r}_{1},\vec{r}_{2},\vec{r}_{3})=\left(
		\begin{array}{c|ccc}
			-\frac{3}{2}\Delta _{a} & g_{1}(\vec{r}_{1}) & g_{2}(\vec{r}_{2}) & g_{3}(\vec{r}_{3}) \\ \hline
			g_{1}(\vec{r}_{1}) & -\frac{1}{2}\Delta _{a} & V_{dd} & V_{dd} \\
			g_{2}(\vec{r}_{2}) & V_{dd} & -\frac{1}{2}\Delta _{a} & V_{dd} \\
			g_{3}(\vec{r}_{3}) & V_{dd} & V_{dd} & -\frac{1}{2}\Delta _{a}%
		\end{array}%
		\right) ,
	\end{equation}
	and the lower-state submatrix can be similarly diagonalized with the unitary matrix in Eq.~(\ref{matrix-Sl-3}). Then the Hamiltonian is transformed into an arrowhead matrix
	\begin{equation}\label{h3-1arrow-real}
		\tilde{H}_{[3]}^{\left( 1\right) }(\vec{r}_{1},\vec{r}_{2},\vec{r}_{3})=\left(
		\begin{array}{c|c}
			\mathbf{U}_{\left[ 3\right] }^{\left( 1\right) } & \mathbf{\tilde{C}}%
			_{\left[ 3\right] }^{\left( 1\right) } \\ \hline
			\left( \mathbf{\tilde{C}}_{\left[ 3\right] }^{\left( 1\right) }\right)
			^{\dag } & \mathbf{\tilde{L}}_{\left[ 3\right] }^{\left( 1\right) }%
		\end{array}%
		\right) ,
	\end{equation}
	where these submatrices $\mathbf{\tilde{L}}_{[3]}^{(1)}$ and $\mathbf{\tilde{C}}_{[3]}^{(1)}$ are given by
	\begin{subequations}\label{3-1-arrow-real}
		\begin{align}
			\mathbf{\tilde{L}}_{[3]}^{(1)} &= \operatorname{diag}\left[ 
			(-\Delta _{a}+4V_{dd})/{2},-(\Delta _{a}+2V_{dd})/{2},-(\Delta _{a}+2V_{dd})/{2}\right] \label{3-1-real-arrow-l},\\
			\mathbf{\tilde{C}}_{[3]}^{(1)} &= [G_{1}(\vec{r}_{1},\vec{r}_{2},\vec{r}_{3}),G_{2}(\vec{r}_{1},\vec{r}_{2},\vec{r}_{3}), G_{3}(\vec{r}_{1},\vec{r}_{2},\vec{r}_{3})]\label{3-1-real-arrow-c},
		\end{align}
	\end{subequations}
	where the coupling strengths are introduced as $G_{1}(\vec{r}_{1},\vec{r}_{2},\vec{r}_{3})=[{g_{1}(\vec{r}_{1})+g_{2}(\vec{r}_{2})+g_{3}(\vec{r}_{3})}]/{\sqrt{3}}$, $G_{2}(\vec{r}_{1},\vec{r}_{2},\vec{r}_{3})=[{-g_{1}(\vec{r}_{1})+g_{2}(\vec{r}_{2})}]/{\sqrt{2}}$, and $G_{3}(\vec{r}_{1},\vec{r}_{2},\vec{r}_{3})=-[{g_{1}(\vec{r}_{1})+g_{2}(\vec{r}_{2})-2g_{3}(\vec{r}_{3})}]/{\sqrt{6}}$. Here, these four new basis states for the Hamiltonian $\tilde{H}_{[3]}^{\left( 1\right) }(\vec{r}_{1},\vec{r}_{2},\vec{r}_{3})$ are given in Eq.~(\ref{four-state}).
	
	The dark states in this case can be obtained by analyzing Eq.~(\ref{h3-1arrow-real}) with the arrowhead-matrix method.
	
	(1) Consider the case of zero coupling column vector:
	(i) When ${g_{1}(\vec{r}_{1})+g_{2}(\vec{r}_{2})+g_{3}(\vec{r}_{3})}=0$, we have $G_{1}(\vec{r}_{1},\vec{r}_{2},\vec{r}_{3})=0$, then the state $\vert L _{\left[3\right] }^{\left( 1\right) }\left(1\right)\rangle$ is decoupled from the upper state $\vert u_{1}\rangle$ and becomes a dark state.
	(ii) When $g_1(\vec{r}_{1})=g_2(\vec{r}_{2})$, the coupling strength $G_{2}(\vec{r}_{1},\vec{r}_{2},\vec{r}_{3})=0$, then the state $\vert L _{\left[3\right] }^{\left( 1\right) }\left(2\right)\rangle$ is decoupled from the upper state $\vert u_{1}\rangle$ and becomes a dark state.
	(iii) When $g_1(\vec{r}_{1})+g_2(\vec{r}_{2})=2g_3(\vec{r}_{3})$, we get $G_{3}(\vec{r}_{1},\vec{r}_{2},\vec{r}_{3})=0$, then the state $\vert L _{\left[3\right] }^{\left( 1\right) }\left(3\right)\rangle$ becomes a dark state.
	
	(2) Consider the case of degenerate lower-state subspace:
	It can be seen from Eq.~(\ref{3-1-real-arrow-l}) that the second and third eigenvalues are identical, then there is a two-dimensional degenerate lower-state subspace $\{\vert L _{\left[3\right] }^{\left( 1\right) }\left(2\right)\rangle,\vert L _{\left[3\right] }^{\left( 1\right) }\left(3\right)\rangle \}$. As a result, there exists one dark state
	\begin{align}
		\vert D_{[3]}^{(1)}\rangle
		&= \frac{1}{\tilde{\mathcal{N}}_{[3]}^{(1)}}
		\big[ G_2(\vec{r}_{1},\vec{r}_{2},\vec{r}_{3})\vert L_{[3]}^{(1)}(3)\rangle
		- G_3(\vec{r}_{1},\vec{r}_{2},\vec{r}_{3})\vert L_{[3]}^{(1)}(2)\rangle \big] \notag \\
		&= \frac{1}{\tilde{\mathcal{N}}_{[3]}^{(1)}}
		\vert 0\rangle \Bigg[
		\left(-\frac{G_2(\vec{r}_{1},\vec{r}_{2},\vec{r}_{3})}{\sqrt{6}}+\frac{G_3(\vec{r}_{1},\vec{r}_{2},\vec{r}_{3})}{\sqrt{2}}\right)\vert e,g,g\rangle \notag \\
		&\quad -\left(\frac{G_2(\vec{r}_{1},\vec{r}_{2},\vec{r}_{3})}{\sqrt{6}}+\frac{G_3(\vec{r}_{1},\vec{r}_{2},\vec{r}_{3})}{\sqrt{2}}\right)\vert g,e,g\rangle\notag \\
		&\quad + \frac{G_2(\vec{r}_{1},\vec{r}_{2},\vec{r}_{3})}{\sqrt{6}}\vert g,g,e\rangle \Bigg],
		\label{3-1dark-real}
	\end{align}
	where the constant $\tilde{\mathcal{N}}_{[3]}^{\left(
		1\right) }=[G_{2}(\vec{r}_{1},\vec{r}_{2},\vec{r}_{3})^{2}+G_{3}(\vec{r}_{1},\vec{r}_{2},\vec{r}_{3})^{2}] ^{1/2}$ is introduced.
		
	The results are in consistent with our previous analyses concerning the dark states in the single-excitation subspace of the three-atom case, and the corresponding dark-state effect can be similarly characterized with the populations of atoms in a proper initial state.
	
	\section{Discussions and conclusion}\label{sec8}
	
	Finally, we present some discussions on the experimental implementation of the system and the experimental observation of the dark-state effect.
	
	(i) In this work, we study the number and form of the dark states in the coupled cavity-Rydberg-atom systems with different numbers of atoms. Therefore, the candidate physical platform to implement our scheme should be able to realize the coupled cavity-Rydberg-atom model. Namely, the candidate setups should be able to realize the Tavis-Cummings couplings between a cavity field and $N$ Rydberg atoms, as well as the dipole-dipole interactions among these atoms. In particular, to control the atom-photon couplings and the dipole-dipole interaction for the appearance of the dark states, the position of the atoms should be chosen on demand. In typical coupled cavity-Rydberg-atom systems, these atoms are controlled by optical tweezers~\cite{ryd-re2}. Due to the development of the experimental techniques for realizing programmable Rydberg-atom array, the control of atom locations should be within the reach of current experimental conditions. In addition, we want to mention that our scheme can also be realized in other physical platforms, in which the Tavis-Cummings interaction between the cavity field and multiple atoms, as well as the dipole-dipole interactions among atoms can be realized. For example, our scheme can be realized in circuit-QED systems, in which both the atom-field and atom-atom interactions can be realized. In particular, these coupling strengths can be designed on demand because the superconducting circuits possess the advantage concerning the controllability and tunability. 
	
	(ii) We also suggest to characterize the dark states by inspecting the populations of some specific quantum states~\cite{circuit-QED1}, which can be detected in experiments. For implementation of the dark state characterization, the system need to be created in some special states by designing proper drivings to either the atoms or the cavity field. The details concerning the initial state preparation have been discussed in the paragraph including Eqs.~(\ref{qudong}). In addition, the dark states can be observed by measuring the atomic populations. As discussed in the above sections concerning the characterization of the dark states, we can distinguish the dark states from the ground state of the system by detecting the excited-state population of these involved atoms.
	
	(iii) About the parameters used in our simulations, we consider the strong-coupling regime for the atom-field couplings. Then we can perform the rotating-wave approximation to reach the Tavis-Cummings interactions. In addition, we take the cavity field dissipation $\kappa/g_1=0.3$ to $0.5$ and the dipole-dipole interaction $V_{dd}/g_1=0.5$. These coherent couplings are larger than or comparable to the decay rates such that the population oscillations can be observed in experiments. These used parameters are consistent with experimentally accessible conditions. All the above discussions indicate that our present scheme is within the reach of state-of-the-art experimental technology.
	                                 
	In conclusion, we have studied the dark-state effect in a coupled cavity-Rydberg-atom system with the arrowhead-matrix method. 
	We have obtained the numbers and form of the dark states in the certain excitation-number subspaces for the two-, three-, four-atom cases in detail. We have also extended the analysis to a general $N$-atom case and obtained the number and form of the dark states in the single-excitation subspace. Based on the number and form of the dark states, we also suggest to characterize the dark state by inspecting the populations of some specific quantum states, which can be detected in experiments.  
	Finally, we have studied the dark-state effect when both the atomic dipole-dipole interaction strengths and the atom-cavity-field coupling strengths depend on the position of the atoms. Based on our theoretical analysis, we have established the existence conditions for dark states in both two- and three-atom cases. In addition, we have performed numerical simulations to substantiate our analytical findings.
	
	\begin{acknowledgments}
		J.-Q.L. was supported in part by the National Natural Science Foundation of China (Grants No. 12247105, No. 12575015, and No. 12421005), National Key Research and Development Program of China (Grant No. 2024YFE0102400), and Hunan Provincial Major Sci-Tech Program (Grant No. 2023ZJ1010). L.M.K. was supported by the NSFC (Grant Nos.~12247105 and~12421005), Quantum Science and Technology-National Science and Technology Major Project (Grant No.~2024ZD0301000), the Hunan Provincial Major Sci-Tech Program (Grant No.~2023ZJ1010),  and the XJ-Lab key project (Grant No.~23XJ02001).
	\end{acknowledgments}

\end{document}